%% file: main.tex
\documentclass[12pt,a4]{article}
\usepackage{graphicx}
\usepackage{amsmath}
\usepackage[utf8]{inputenc}
\usepackage[T1]{fontenc}
\usepackage{mathptmx}
\usepackage{xcolor}
\definecolor{myblue}{rgb}{0.0,0.0,0.0}
\usepackage{draftwatermark}
\usepackage{amssymb}                                  
\usepackage{cite}
\usepackage{tabularx}                                 
\usepackage{hyperref}                                 
\usepackage[all]{hypcap}                              
\usepackage{tikz} 
\usepackage{lipsum}
\usepackage{fancyhdr}
\fancyheadoffset{0.02\textwidth}
\fancyfootoffset{0.02\textwidth}

\newlength{\dinwidth}
\newlength{\dinmargin}
\def\ee{e^+e^-}
\newcommand\pubnumber{ILC-NOTE-2015-068\\
DESY 15-102\\
IHEP-AC-2015-002\\
KEK Preprint 2015-17\\
SLAC-PUB-16309}
\newcommand\pubdate{June 25, 2015}
\newcommand\pubblock{\rightline{\begin{tabular}{l} \pubnumber\\
         \pubdate \end{tabular}}}

\begin{document}

\begin{titlepage}
\begin{flushleft}
\end{flushleft}

\pubblock

\vspace{1.0cm}

\begin{center}
\begin{Large}

 {\bfseries \boldmath ILC Operating Scenarios}

\vspace{1.5cm}

ILC Parameters Joint Working Group

\end{Large}

T. Barklow, J. Brau, K. Fujii, J. Gao, J. List,
N. Walker, K. Yokoya


\end{center}

\vspace{1cm}

\begin{abstract}
The ILC Technical Design Report documents the design for the construction of
   a linear collider which can be operated at energies up to
   500 GeV.  
   This report summarizes the outcome of a study of
   possible running scenarios,
   including a realistic estimate of the real time
   accumulation of integrated luminosity based on ramp-up
   and upgrade processes.  
   The evolution of the physics outcomes is
   emphasized, including running initially at 500 GeV, then at 350 GeV and 250 GeV.
   The running scenarios have been chosen 
  to optimize the Higgs precision measurements and top physics while searching for 
  evidence for signals beyond the standard model, including dark matter.
   In addition to the certain precision physics on the Higgs and top that is the main focus of this
   study, there are scientific motivations that indicate the
   possibility for discoveries of new particles 
   in the upcoming operations of
   the LHC or the early operation of the ILC.  Follow-up studies of such discoveries 
   could alter the plan for the
   centre-of-mass collision energy of the ILC and expand 
   the scientific impact of the ILC physics program.
   It is envisioned that a decision on a possible energy upgrade
   would be taken near the end of the twenty year period considered in this report.
\end{abstract}


\vspace{1.0cm}

\begin{center}
\end{center}

\end{titlepage}
\input{intro.tex}


\section{ILC500 Runnning Scenarios}
In this section, we introduce the energy steps, integrated luminosities and polarisation
sharing of the studied running scenarios and discuss the considerations which lead to
our choices.
\label{sec:runscen}
\subsection{Integrated luminosities and polarisation splitting}
\label{sec:lumipol}
\input{lumipol.tex}

\subsection{Operation Scenarios}
\label{sec:opscen}
\input{scenarios.tex}
\input{realtime.tex}

%
\input{physics.tex}

%
\input{tth.tex}
%

\input{additional.tex}

\input{conclusions.tex}

\section*{Acknowledgement}
Many members of the ILC community contributed to this report by valuable discussions as well 
as sharing of results and code. In particular we'd like to thank Mikael Berggren, Roberto Contino, Christophe Grojean, Benno List, Maxim Perelstein, Michael Peskin, 
Roman P\"oschl, Juergen Reuter, Tomohiko Tanabe, Mark Thomson, Junping Tian, 
Graham Wilson and all members of the LCC Physics Working Group.

\clearpage

\input{higgsMeasurements.tex}

\input{Appendix-projected.tex}

\clearpage

\input{bib.tex}


\end{document}

%% file: intro.tex
\section{Introduction}
\label{sec:intro}
The ILC requirements document ``Parameters for the Linear Collider"~\cite{Heuer:2006} describes the basic requirements for a 500 GeV centre-of-mass  machine with the possibility of extending the energy up to 1 TeV.  The  ILC design given in the Technical Design Report (TDR)~\cite{Adolphsen:2013kya} realizes this machine. 

Following the considerations given to the machine construction issues, a study has been conducted to understand the physics implications of the choice
of the sequence of operating energies. The ILC Parameters Joint Working Group, created by the LCC Directorate,
was charged with this study.

In order to quantify the impact of various options of running on the physics output, and particularly on its evolution with time, a number of operating scenarios were  
studied. 
Operations may be conducted at the full centre-of-mass energy
of the ILC (500 GeV) or at a lower energy than the full capability of the collider.

It must be emphasized that one of the very valuable features of the ILC is its energy flexibility.
Discoveries made at the LHC or in the early phases of the ILC will impact the choice of operating 
energies; even though the collider has the capability
of a particular maximum energy reach, it can operate at lower
energies as the physics requirements dictate.
Such potential discoveries cannot be specified with certainty today, but would greatly expand
   the scientific impact of the ILC and broaden its follow-on program.

The principal physics motivations for operations in the $250$-$500$\,GeV range are a high-precision characterisation 
of the recently discovered Higgs boson and high-precision determination of the properties of the top quark. Together with precision measurements of the $W$ and $Z$ boson masses and couplings, they provide powerful tools for discovering beyond the Standard Model physics. The precision part of the ILC program will be complemented by searches for direct production of new particles, in particular for Dark Matter candidates. Once new particles in the kinematic reach of the ILC have been discovered, either at the LHC or at the ILC itself, precision measurements of their masses and chiral couplings will unveil the type of underlying extension of the Standard Model (SM) and its fundamental parameters. A broad review of the physics program of the ILC can be found in the Physics Volume of the TDR~\cite{Baer:2013}, while a comprehensive and updated summary is available in~\cite{LCCPhysGroup:2015}.

Before comparing different operating scenarios for the ILC, we briefly summarize the interplay 
of the various center-of-mass energies for the main topics of the ILC physics program:

\begin{itemize}

\item{Higgs precision measurements:} 
In $e^+e^-$ collisions, Higgs bosons can be produced either via their coupling to the $Z$ boson (``Higgsstrahlung''), 
or via their coupling to the $W$ boson (``$WW$ fusion''). At $\sqrt{s} = 250$\,GeV, the production occurs dominantly through Higgsstrahlung. This is very beneficial for the fully model-independent recoil method for measuring
the total $ZH$ cross section and the Higgs mass, but on the other hand there is limited sensitivity to the
Higgs-$W$ coupling. For $350$\,GeV $\lesssim \sqrt{s} \lesssim 550$\,GeV, these two production modes are of similar size, which yields a balanced sensitivity to both the Higgs-$W$ and the Higgs-$Z$ couplings. Probing the top Yukawa coupling from
associated production of $t\bar{t}H$ requires at least $\sqrt{s} \gtrsim 500$\,GeV. Due to these three production processes, Higgs physics exhibits the most complex interplay between different center-of-mass energies, which we will
discuss in more detail in the following sections.

\item{Top quark precision measurements:} 
At the ILC with  $\sqrt{s} \gtrsim 350$\,GeV, top quark pairs will be produced for the very first time in $e^+e^-$ 
collisions. A variation of $\sqrt{s}$ near $350$\,GeV will give a quantitatively and qualitatively unique measurement
of the top quark mass. However, deviations of the electroweak couplings of the top quark from their SM values, which are predicted by many extensions of the Standard Model, can only be detected at higher $\sqrt{s} \gtrsim 450$\,GeV, since near threshold they are buried beneath the remnant of the strong $t\bar{t}$ resonance.

\item{Couplings and masses of the $W$ and $Z$ bosons:}
A powerful probe for physics beyond the Standard Model are triple and quartic gauge couplings. 
While they can in principle be probed at any center-of-mass energy which allows di-boson production, the sensitivity
rises strongly with $\sqrt{s}$. For instance in the case of charged triple gauge couplings, the ILC with $\sqrt{s} = 500$\,GeV will increase the sensitivity by two orders of magnitude with respect to LEP2. The $W$ boson mass measurement 
can reach the few MeV regime either from a scan of the production threshold near $\sqrt{s} = 161$\,GeV, or from
kinematic reconstruction at $\sqrt{s} = 500$\,GeV. Concerning the $Z$ boson, operation of the ILC near $\sqrt{s} = 90$\,GeV with polarised beams could be the only possibility to resolve the existing tension between the left-right
asymmetry measured at SLD and the forward-backward asymmetry from LEP~\cite{ALEPH:2005ab}. 
Also an improvement of the $Z$ mass measurement
beyond LEP is conceivable{\color{myblue}~\cite{Wilson:BeamEnergy}}.  These lower energy operations ($\sqrt{s}<200$\,GeV) would require 
machine upgrades from the TDR design for optimal luminosities.

\item{Production of new particles, including Dark Matter:}
The reach for the pair production of new particles and thus the possibility for direct discoveries obviously increases
with center-of-mass energy and covers almost $M <  \sqrt{s}/2$. In new physics models where the production cross sections
are not, like in supersymmetry, given by the SM couplings, higher center-of-mass energy allows the probing of smaller couplings, thus higher scales of new physics.

\end{itemize}

These physics motivations have driven the design of the operating scenarios.

\begin{figure}[ht]
\centering
\includegraphics[width=0.9\textwidth]{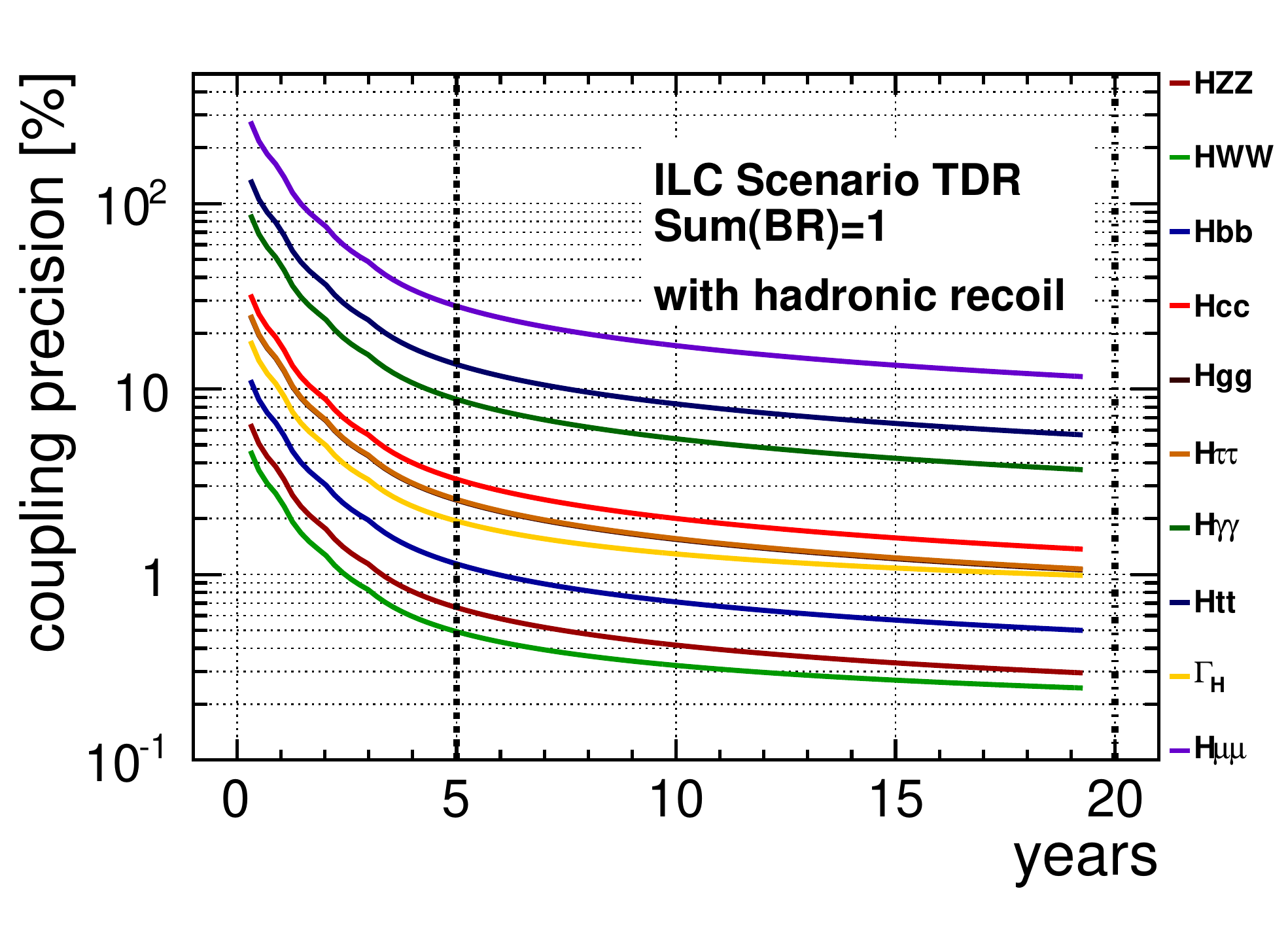}
  \caption{{Statistical uncertainty of Higgs couplings in the TDR baseline ILC operating at $\sqrt{s}=500$\,GeV as a function of real time, including ramp-up as discussed in section~\ref{sec:realtime}, and assuming the luminosity sharing between the beam helicity configurations discussed in section~\ref{sec:lumipol}. See section~\ref{subsec:physics_higgs}
for definition of the $\sum_{i} {BR_i}=1$ constraint. } \label{fig:Higgs500}}
\end{figure}

The TDR presented a baseline machine with $\sqrt{s}=500$\,GeV. Figure~\ref{fig:Higgs500} shows the time evolution
of the statistical precisions on the Higgs' couplings achievable with this baseline machine
operating at $\sqrt{s}$=500 GeV
 as a function of real 
time until a total integrated luminosity of $2.5$\,ab$^{-1}$ is reached. This time evolution serves as a reference 
for comparison with different scenarios.


The remainder of this report is structured as follows: 
We start in section~\ref{sec:runscen} by introducing some scenarios for operating
the ILC which have been studied and are reported here. 
In section~\ref{sec:realtime} gives an explanation and presentation
of the derivation of the real time needs of the scenarios. Section~\ref{sec:physics} gives examples for the time evolution of different physics observables in selected running scenarios. 
Section~\ref{sec:tth} discusses the top Yukawa coupling reach at the maximum energy of $\sqrt{s}\sim 500$\, GeV. In section~\ref{sec:modNP}, we discuss possible modifications of the
running scenarios in case new phenomena appear at the LHC or the initial runs of the ILC itself.
Finally we comment on the luminosity needs at other centre-of-mass energies in section~\ref{sec:other}, including both operation below $\sqrt{s}=200$\,GeV
and at $\sqrt{s}=1$\,TeV, before concluding the report in section~\ref{sec:concl}.

%% file: lumipol.tex

The total integrated luminosities collected at various center-of-mass
energies will determine the ultimate physics reach of the ILC.
It is not completely clear now what the best combination of dataset 
sizes will be following the upcoming LHC running as well as early ILC operation.
We have studied scenarios which could be chosen based on what is known of the
physics at that time.  {\color{myblue} We compare three scenarios, G-20, H-20 and I-20, each 
involving approximately 20 years of operating the 500 GeV collider, including
shutdowns, a luminosity upgrade~\cite{Harrison:2013nva} after 8-10 years
and operation efficiencies,
 as will be detailed in section~\ref{sec:realtime}.}
These operating scenarios illustrate different balancing 
between $\sqrt{s}=250$, $350$, and $500$\,GeV and their 
total accumulated luminosities for each energy is presented
in table~\ref{tab:lumiabstot}.
{\color{myblue} 
For comparison, we also list a scenario ``Snow'' which is based on the integrated
luminosities proposed in the ILC Higgs Whitepaper for Snowmass~\cite{Asner:2013psa}, plus a $200$\,fb$^{-1}$
at the top threshold.}  It must be kept in mind that other operating energies may be required based on
the physics results from the LHC or early ILC.

%
%
%
%
%
%
%
%
%

\begin{table}[h]
\centering
  \renewcommand{\arraystretch}{1.10}
\begin{tabularx}{\textwidth}{*{4}{>{\centering\arraybackslash}X} || *{1}{>{\centering\arraybackslash}X}} 
\hline
            &  \multicolumn{4}{c}{$\int{\mathcal{L} dt}$ [fb$^{-1}$]} \\
\hline
$\sqrt{s}$  & G-20      &   H-20   &  I-20   & Snow   \\
\hline
250\,GeV    &  500      &  2000    &   500   & 1150   \\
350\,GeV    &  200      &   200    &  1700   &  200  \\
500\,GeV    & 5000      &  4000    &  4000   & 1600  \\
\hline
\end{tabularx}
\caption{Proposed total target integrated luminosities for $\sqrt{s}=250$,  $350$, $500$\,GeV {\color{myblue},
based on $20$ ``real-time'' years of ILC operation under scenarios G-20, H-20 and I-20. The total integrated luminosities assumed for Snowmass
are listed for comparison based on 13.7 ``real-time'' years.}}
\label{tab:lumiabstot} 
\end{table}

The ultimate physics reach further depends on the assumed beam polarisations. Concerning the
absolute values, the highest achievable degree of polarisation is desirable, in particular for 
the positron beam. We assume here the TDR values of $|P(e^-)|=80\%$ and $|P(e^+)|=30\%$,
although higher values are possible for both species. 
The option of upgrading to $|P(e^+)|=60\%$ would enhance the physics potential of the machine, 
while the 
absence of positron polarisation would reduce it. 
The size of the impact depends 
strongly on the individual physics observables and is beyond the scope of this report, but should
be quantified in the near future.

Independent of the absolute values of the beam polarisations, the choice of their signs, i.e. 
running with predominantly left- or right-handed electrons / positrons is important for the physics
program. Due to conservation of angular momentum, the $s$-channel exchange of $Z$ bosons or 
photons is only possible for opposite-sign chirality, i.e. $e^-_Le^+_R$ or $e^-_Re^+_L$. Among these, the chiral couplings of the $Z$ boson prefer the former combination. Within the SM, only the $t$-channel exchange of a $Z$ boson or photon is allowed for like-sign helicities, i.e. $e^-_Le^+_L$ and $e^-_Re^+_R$. 
Beyond the SM, Majorana particles, for instance neutralinos in supersymmetric extensions, could be exchanged in the $t$-channel,
allowing like-sign helicities as eg in selectron production.
Only the combination $e^-_Le^+_R$ contributes to $t$-channel exchange of a $W$ boson or a electron-neutrino. Thus $W$ pair production, which is an important background for many signatures,
can be reduced by orders of magnitude by choosing the opposite beam helicities.  

\begin{table}[h]
\centering
  \renewcommand{\arraystretch}{1.10}
\begin{tabularx}{\textwidth}{*{5}{>{\centering\arraybackslash}X}} 
\hline
        & \multicolumn{4}{c}{fraction with $\operatorname{sgn}(P(e^-),P(e^+))= $ } \\
           & (-,+) & (+,-) & (-,-) & (+,+) \\
\hline
$\sqrt{s}$ & [\%]  &  [\%] & [\%]  & [\%]  \\ 
\hline
250\,GeV   & 67.5 &  22.5 &  5    &   5   \\
350\,GeV   & 67.5 &  22.5 &  5    &   5   \\
500\,GeV   &  40  &  40   &  10   &  10   \\
\hline
\end{tabularx}
\caption{Relative sharing between beam helicity configurations proposed for the various center-of-mass energies.}
\label{tab:pollumirel} 
\end{table}

Therefore we propose the sharing between the four possible sign combinations listed in table~\ref{tab:pollumirel}. It should be noted that one profits from the cancellation of experimental systematic uncertainties between these samples only if they are accumulated 
while flipping the beam helicities in a randomized way on bunch train time-scales. 
The average flipping frequencies are thereby adjusted to give the helicity fractions listed in table~\ref{tab:pollumirel}.

At $\sqrt{s}=250$ and $350$\,GeV, it is expected that the main interest will be on Standard Model processes. Thus $90\%$ of the data is collected in the unlike-sign combinations, preferring left-handed electrons over right-handed. Only a small fraction ($10\%$) of the like-sign configurations is planned 
to control systematics.

\begin{figure}[ht]
\centering
\includegraphics[width=0.9\textwidth]{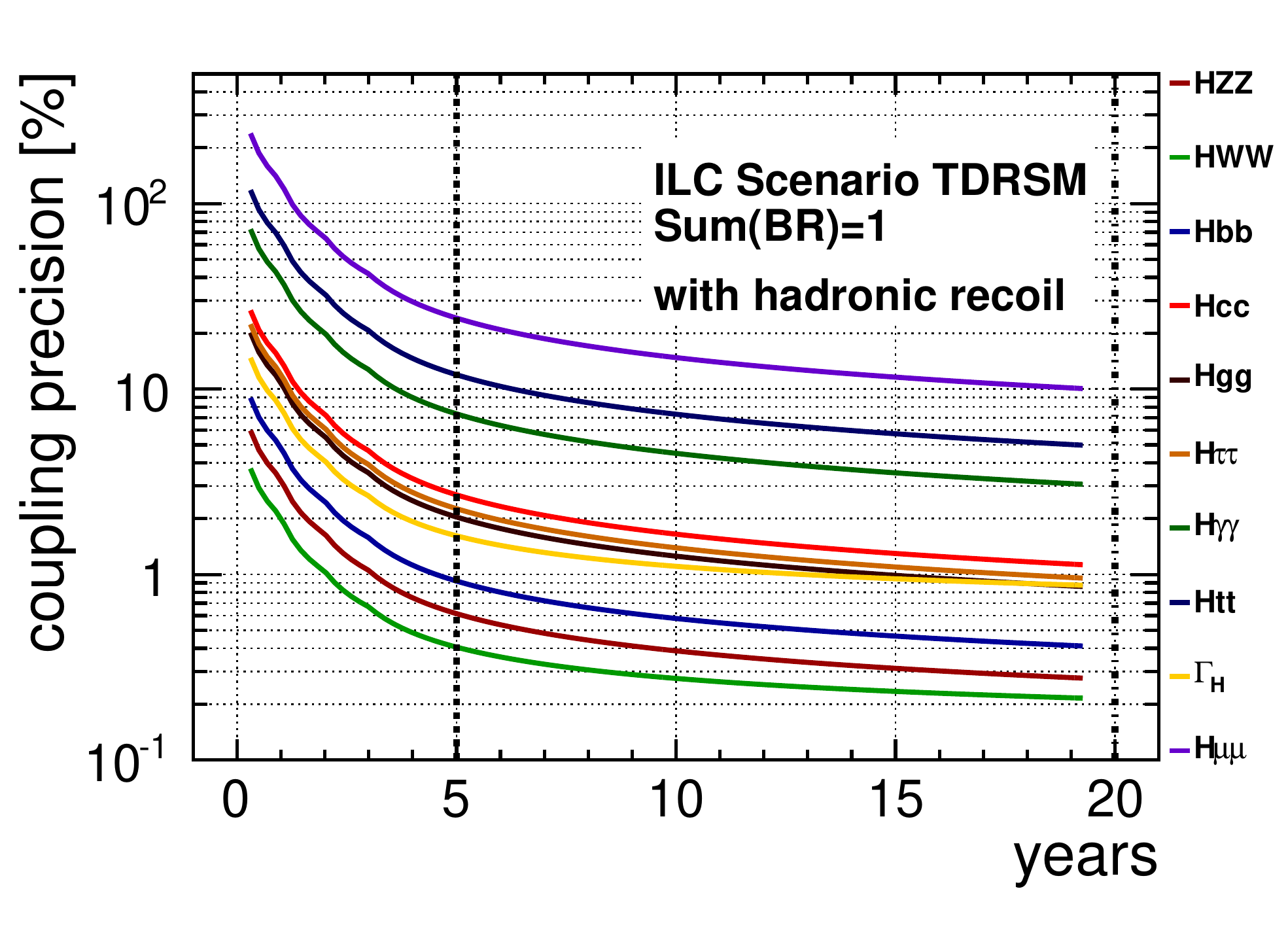}
  \caption{{Statistical uncertainty of Higgs couplings in the TDR baseline ILC operating at $\sqrt{s}=500$\,GeV as a function of real time, including ramp-up as discussed in section~\ref{sec:realtime}, with a luminosity splitting between beam helicities of (67.5\%,22.5\%,5\%,5\%).} \label{fig:Higgs500_LR675}}
\end{figure}

At higher $\sqrt{s}$, the picture changes because new physics is more likely to appear. 
For example Dark Matter searches with an effective field theory approach could profit significantly from like-sign data-taking depending on the Lorentz structure of its couplings;
also the determination of the chiral properties of new particles requires
a more balanced sharing between beam helicity configurations. Indirect searches, e.g. via
the electroweak couplings of the top quark, prefer right-handed electrons over left-handed ones.
Thus, a splitting of (40\%,40\%,10\%,10\%) is proposed here. On the other hand, Higgs production
via $WW$ fusion exists only for left-handed electrons and right-handed positrons. Figure~\ref{fig:Higgs500_LR675}
shows the same time evolution of precisions as figure~\ref{fig:Higgs500}, but with a splitting of
(67.5\%,22.5\%,5\%,5\%). 

The largest improvement occurs as expected for $g_{HWW}$, which after $5$ years reaches a precision of $\sim0.4\%$,
to be compared to $\sim0.5\%$ with (40\%,40\%,10\%,10\%) (c.f. fig~\ref{fig:Higgs500}). We conclude
that this improvement does not outweigh the expected loss of sensitivity in other physics,
 e.g. Dark Matter searches
or measurement of the top quark couplings. Thus, we apply the helicity sharing listed in table~\ref{tab:pollumirel}
for all scenarios.

\begin{table}[h]
\centering
  \renewcommand{\arraystretch}{1.10}
\begin{tabularx}{\textwidth}{*{5}{>{\centering\arraybackslash}X}}    
\hline
        &  \multicolumn{4}{c}{integrated luminosity with $\operatorname{sgn}(P(e^-),P(e^+))= $ } \\
           & (-,+)       & (+,-)       & (-,-)       &  (+,+)     \\
\hline
$\sqrt{s}$ & [fb$^{-1}$] & [fb$^{-1}$] &  [fb$^{-1}$] & [fb$^{-1}$] \\ 
\hline
250\,GeV    &  1350      &  450        &  100	      &   100  \\
350\,GeV    &   135      &   45	       &   10	      &    10  \\
500\,GeV    &  1600      & 1600        &  400	      &   400  \\
\hline
\end{tabularx}
\caption{Integrated luminosities per beam helicity configuration resulting from the fractions in table~\ref{tab:pollumirel} in scenario H-20.
}
\label{tab:pollumiabs} 
\end{table}

Table~\ref{tab:pollumiabs} shows an example case of the resulting integrated luminosities per center-of-mass energy and helicity configuration for the scenario H-20.

It must be stressed once more that a key asset of the ILC is its flexibility.  For all center-of-mass energies,
further discoveries at the LHC or the results of the first ILC runs could lead to modifications of the ideal sharing
 between helicity fractions.
 Such changes in the run plan can easily be accommodated based on future physics results.

%

%% file: scenarios.tex

The total integrated luminosities presented in section~\ref{sec:lumipol} 
are collected at different stages of the machine in different periods of time, 
leading to what we refer to as ``running scenarios''. 
In this section, we propose a few examples of such running scenarios to be
evaluated from the physics perspective. 

We concentrate on two main parameters to vary:

{\color{myblue}

\begin{itemize}
  \item The time before the luminosity upgrade~\cite{Harrison:2013nva}: Scenarios H-20 and I-20 foresee
   the luminosity upgrade after approximately $8$ years, while scenario G-20
   assumes the luminosity upgrade later, only after accumulating two more years
   of integrated luminosity at $500$\,GeV, after $10$ years. 
  \item The final accumulation of integrated luminosity per energy: Scenario G-20 includes only small data sets
        at $250$ and $350$\,GeV and focusses on collecting the largest possible luminosity at the top baseline 
        energy. In contrast, scenarios H-20 and I-20 illustrate the effect of taking a large dataset at $250$\,GeV
        or $350$\,GeV, respectively.
 \end{itemize}

}

In this we apply the following guidelines/restrictions:

\begin{itemize}
  \item All scenarios are limited to about equal total operation times near $20$ years, before 
      a possible 1\,TeV upgrade or other running options. 
  \item All scenarios assume that an integrated luminosity of at least $200$\,fb$^{-1}$ will be 
       collected at the top threshold
      near $\sqrt{s}=350$\,GeV. Current studies~\cite{FrankSimon} 
      indicate that the top mass measurement from a scan of the production threshold becomes theoretically
      limited for $100$\,fb$^{-1}$. In order to allow for data taking with other polarisations, collection
      of control samples and further improvements of the theoretical calculations we assume here a minimum
      of  $200$\,fb$^{-1}$.
      We assume that this will be done with the 500\,GeV machine
       operated at a reduced gradient, after an initial exploration of the physics landscape at 
       $\sqrt{s}=500$\,GeV and with a well run-in machine. 
  \item At each $\sqrt{s}$, the total integrated luminosities given below should be understood 
      to be split up between
      the four possible beam helicity configurations as specified in section~\ref{sec:lumipol}.
  \item In order to give a complete picture of the potential of the ILC, we include a luminosity 
        upgrade and the possibility to provide collisions at more than $5$\,Hz, when surplus
        RF and cryogenic power allows.  
  \item The exact details of the very long term program will 
      depend on future developments at early stages of the ILC, at the LHC and possibly  
       other scientific results. Thus
      we do not speculate here about the possible variations beyond a $\sim 20$ year program, 
      in particular the upgrade to $\sqrt{s}=1$\,TeV. However we note
      that some measurements of the $500$\,GeV program could be done even better at 
      $\sqrt{s}=1$\,TeV. Thus a realisation of the $1$-TeV-upgrade could replace part of the
      required luminosity at $\sqrt{s}=500$\,GeV.
  \item Further discoveries, in particular also at the inital run 
       of ILC500, will change the details
      and might add the necessity to run at additional intermediate energies, 
      either for scanning production thresholds
      of new particles, or for disentangling several states close by in mass.
      We discuss possible modifications of the proposed scenarios in view
      of discoveries in section~\ref{sec:modNP}.  
  \item We do not list here physics running at the $Z$-pole or at the $WW$-threshold. 
      However we note that their physics
      program should be pursued at some point, where the timing will depend on the outcome of an 
      inital running at $\sqrt{s}=500$\,GeV. Optimal luminosities at these
      energies ($\sqrt{s}<200$\,GeV) will require machine upgrades from the TDR design.
  \item Runtime on the $Z$-pole for calibration is also not included. This may be needed annually for
      detector calibration (e.g. of the momentum scale of the tracking detectors). 
      \emph{Here, more precise specifications
      from the experiments are needed in order to assess the amount of data needed for 
      reaching which level of calibration precision.}        
  \item The details about the time lines for these scenarios including ramp-up 
      and upgrade-installation times will be presented in section~\ref{sec:realtime}.     
\end{itemize}

Table~\ref{tab:runscen} shows the integrated luminosities and real time required for each stage of the 
running scenarios presented here. The time estimates include ramp-up efficiencies and installation times
for upgrades. Table~\ref{tab:runtime} summarizes the run times of the scenarios defined in
table~\ref{tab:runscen}.

\begin{table}[h]
\centering
\def\tabularxcolumn#1{m{#1}}
\renewcommand{\arraystretch}{1.0}
\begin{tabularx}{\textwidth}{c | c | *{3}{>{\centering\arraybackslash}X} |  *{3}{>{\centering\arraybackslash}X} } 
\hline
         &Stage                            &   \multicolumn{3}{c|}{500}& \multicolumn{3}{c}{500 LumiUP}   \\
\hline
Scenario&$\sqrt{s}$ [GeV]                  &      500  &  350  &   250 &  500   &   350  &   250    \\ 
\hline 
G-20  &$\int{\mathcal{L} dt}$ [fb$^{-1}$]  &     1000  &  200  &   500 &  4000  &    -   &    -     \\
      &time [years]                        &      5.5  &  1.3  &   3.1 &   8.3  &    -   &    -     \\
\hline
H-20 & $\int{\mathcal{L} dt}$ [fb$^{-1}$]  &      500  &  200  &   500 &  3500  &    -   &   1500   \\
     &time [years]                         &      3.7  &  1.3  &   3.1 &   7.5  &    -   &    3.1   \\ 
\hline
I-20& $\int{\mathcal{L} dt}$ [fb$^{-1}$]   &      500  &  200  &   500 &  3500  & 1500   &     -    \\
      &time [years]                        &      3.7  &  1.3  &   3.1 &   7.5  &  3.4   &     -    \\ 
\hline
\hline
         &Stage                            &   \multicolumn{3}{c|}{500}& \multicolumn{3}{c}{500 LumiUP}   \\
\hline
Scenario&$\sqrt{s}$ [GeV]                  &      250  &  500  &   350 &  250   &   350  &   500    \\ 
\hline  
Snow & $\int{\mathcal{L} dt}$ [fb$^{-1}$]  &      250  &  500  &   200 &  900   &    -   &  1100    \\
     &time [years]                         &      4.1  &  1.8  &   1.3 &   3.3  &    -   &   1.9    \\ 
\hline
\end{tabularx}
\caption{Final integrated luminosities and real time (calendar years) required for each stage of the running scenarios, including ramp up and installation times for upgrades. Not included: calibration and
physics runs at $Z$ pole and $WW$-threshold, scanning of new physics thresholds.
The order of centre-of-mass energies for each scenario correspond to the sequence 
of operations for that scenario.
The ``Snow" scenario results in lower integrated luminosity due to the
shorter assumed ``real-time" of 13.7 years.
}
\label{tab:runscen} 
\end{table}

The motivations for each of these scenarios are as follows:
\begin{itemize}
\item Scenario G-20 emphasizes the data-taking at the top baseline energy. It starts with
      an initial run at $\sqrt{s}=500$\,GeV collecting $1$\,ab$^{-1}$, which is beneficial for
      early results on top electroweak couplings, the top Yukawa coupling, double Higgs production
      as well as for searches. This is followed by rather short dedicated runs at the top threshold and the 
      Higgsstrahlung cross section maximum. After the luminosity upgrade, a very high-statistics
      dataset is collected at $500$\,GeV. {\color{myblue} This will result in a better final performance
      for all measurements which can only be carried out at $\sqrt{s} \ge 500$\,GeV, in particular the
      top Yukawa coupling and the Higgs self-coupling. However this
      scenario fully relies on the hadronic recoil method to deliver sufficiently model-independent
      access to the $Z$-Higgs coupling {\itshape and} on the kinematic reconstruction of $H\to b\bar{b}$
      and $H\to WW^*$ decays to enable a sufficiently precise measurement of the Higgs mass.} 
\item Scenarios H-20 and I-20 have a slightly reduced amount of data at $500$\,GeV, which is 
      complemented by substantial datasets at $250$ and $350$\,GeV, respectively. In both cases,
      the initial run at $\sqrt{s}=500$\,GeV is shortened w.r.t. G-20, allowing for an earlier luminosity
      upgrade. This in turn enables the collection of large datasets at $250$ (H-20) or $350$\,GeV (I-20)
      with only a moderate loss of integrated luminosity at $\sqrt{s}=500$\,GeV. {\color{myblue} Especially
      scenario H-20 with its substantial amount of data collected at $\sqrt{s}=250$\,GeV guarantees the
      fully model-independent profiling of the Higgs boson.}
\item The scenario ``Snow'' follows the scenario developed by the authors of the ILC Higgs Whitepaper 
      for the Snowmass Community Study~\cite{Asner:2013psa} {\color{myblue}  in terms of the time ordering of the data-taking
      at diffrerent center-of-energies and in terms of total integrated luminosities. However, a run at the
      $t\bar{t}$ production threshold has been added. This scenario serves here for comparison purposes. }  
      
\end{itemize}

\begin{table}[h]
\centering
  \renewcommand{\arraystretch}{1.10}
\begin{tabularx}{\textwidth}{*{3}{|>{\centering\arraybackslash}X|}} 
\hline
          &  \multicolumn{2}{c|}{total run time {\itshape before}}  \\
\hline
          &  Lumi upgrade & potential TeV upgrade    \\
\hline
Scenario  & [years]       &  [years]       \\ 
\hline
G-20      &    9.8        &  19.7         \\
H-20      &    8.1        &  20.2         \\
I-20      &    8.1        &  20.4         \\
Snow      &    7.1        &  13.7         \\
\hline
\end{tabularx}
\caption{Cumulative running times for the four scenarios, including ramp-up and installation of upgrades.  
Not included: calibration and
physics runs at $Z$ pole and $WW$-threshold or scanning of new physics thresholds.}
\label{tab:runtime} 
\end{table}

%% file: realtime.tex
\section{Timelines of the running scenarios}
\label{sec:realtime}

\begin{figure}[t]
\centering
\includegraphics[width=0.85\textwidth]{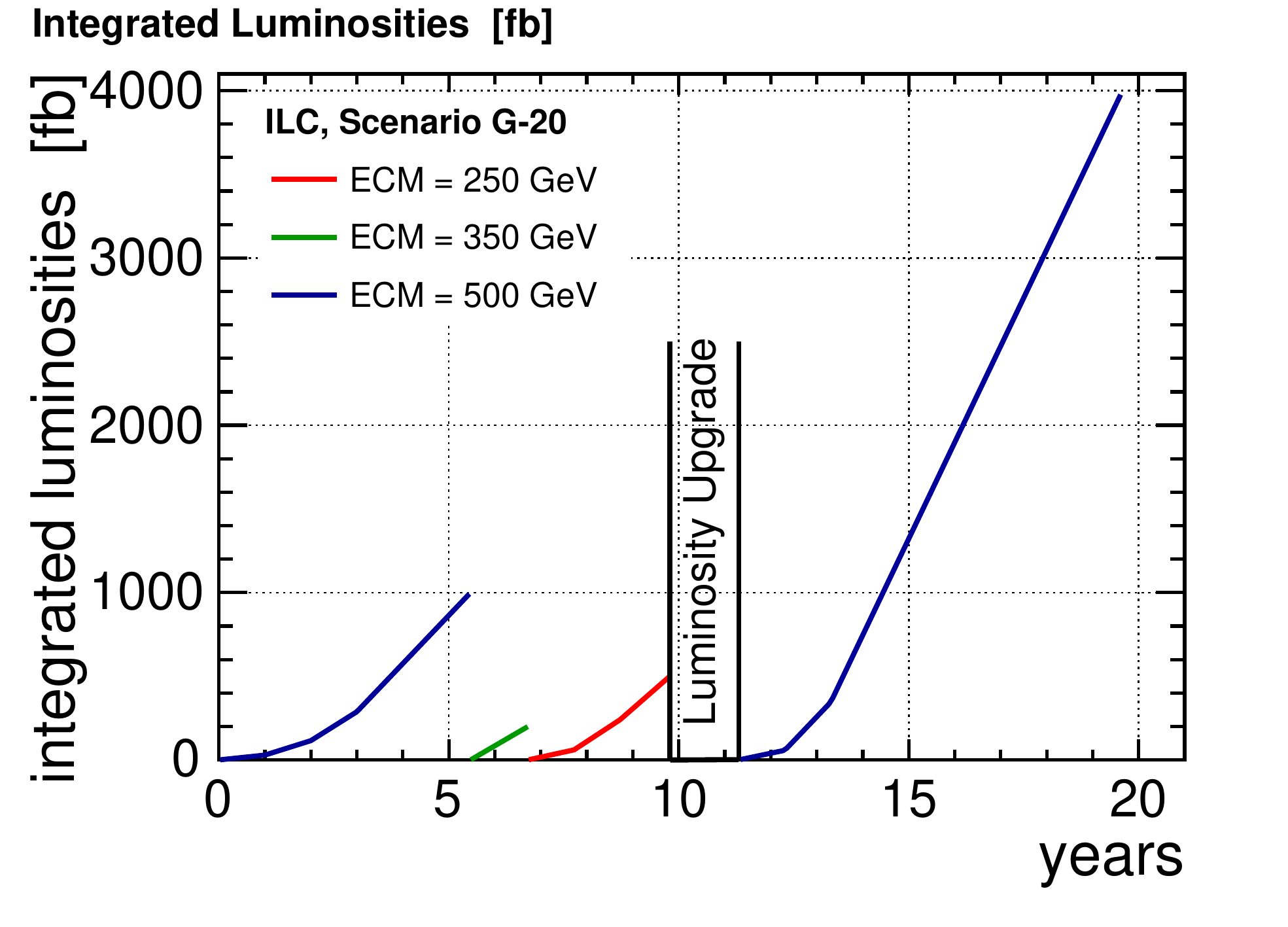}
  \caption{Accumulation of integrated luminosity versus real time for scenario G-20.
  \label{fig:realtime-G-20}}
\end{figure}

\begin{figure}[b]
\centering
\includegraphics[width=0.85\textwidth]{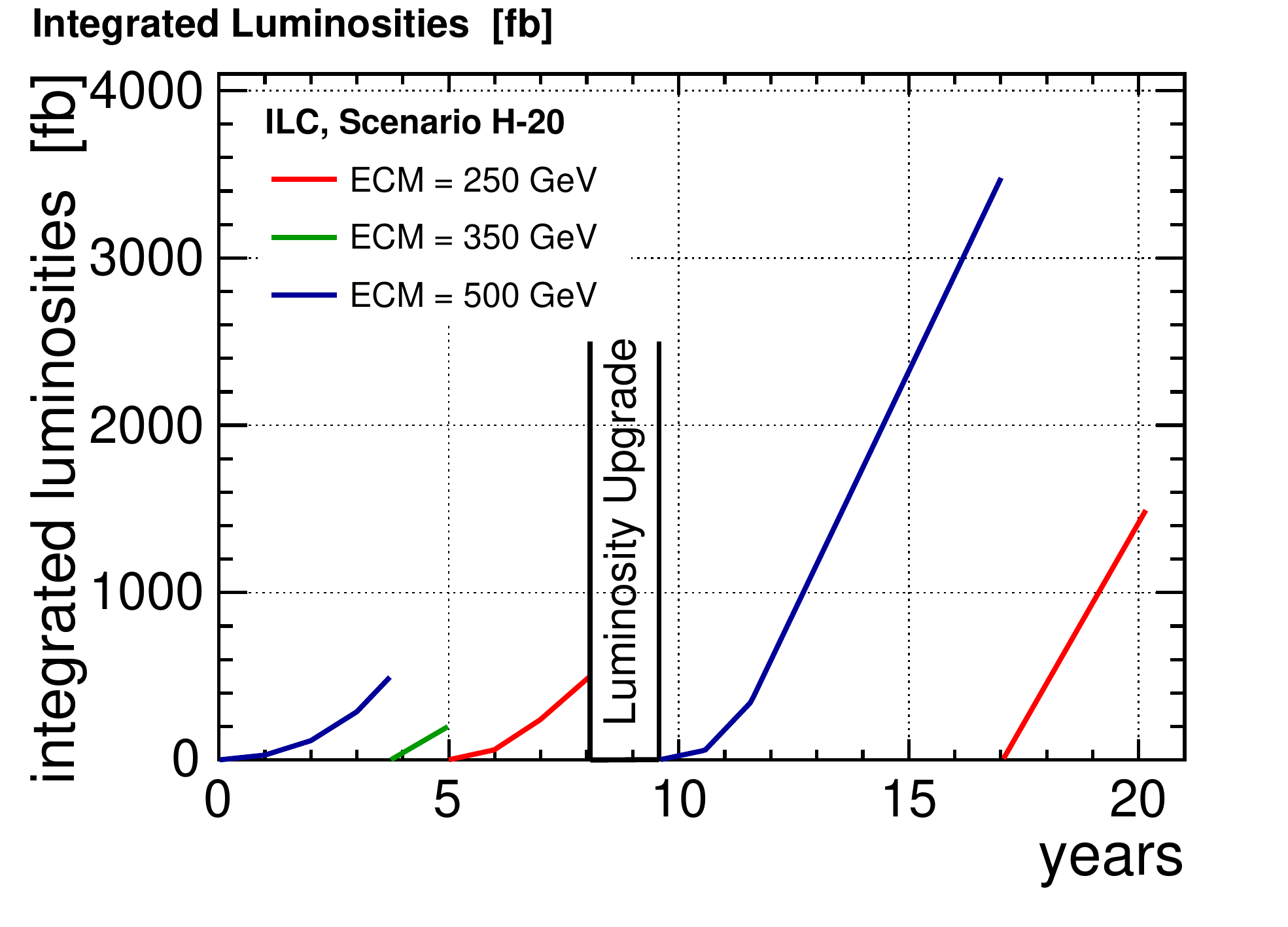}
  \caption{Accumulation of integrated luminosity versus real time for scenario H-20.
  \label{fig:realtime-H-20}}
\end{figure}

\begin{figure}[t]
\centering
\includegraphics[width=0.85\textwidth]{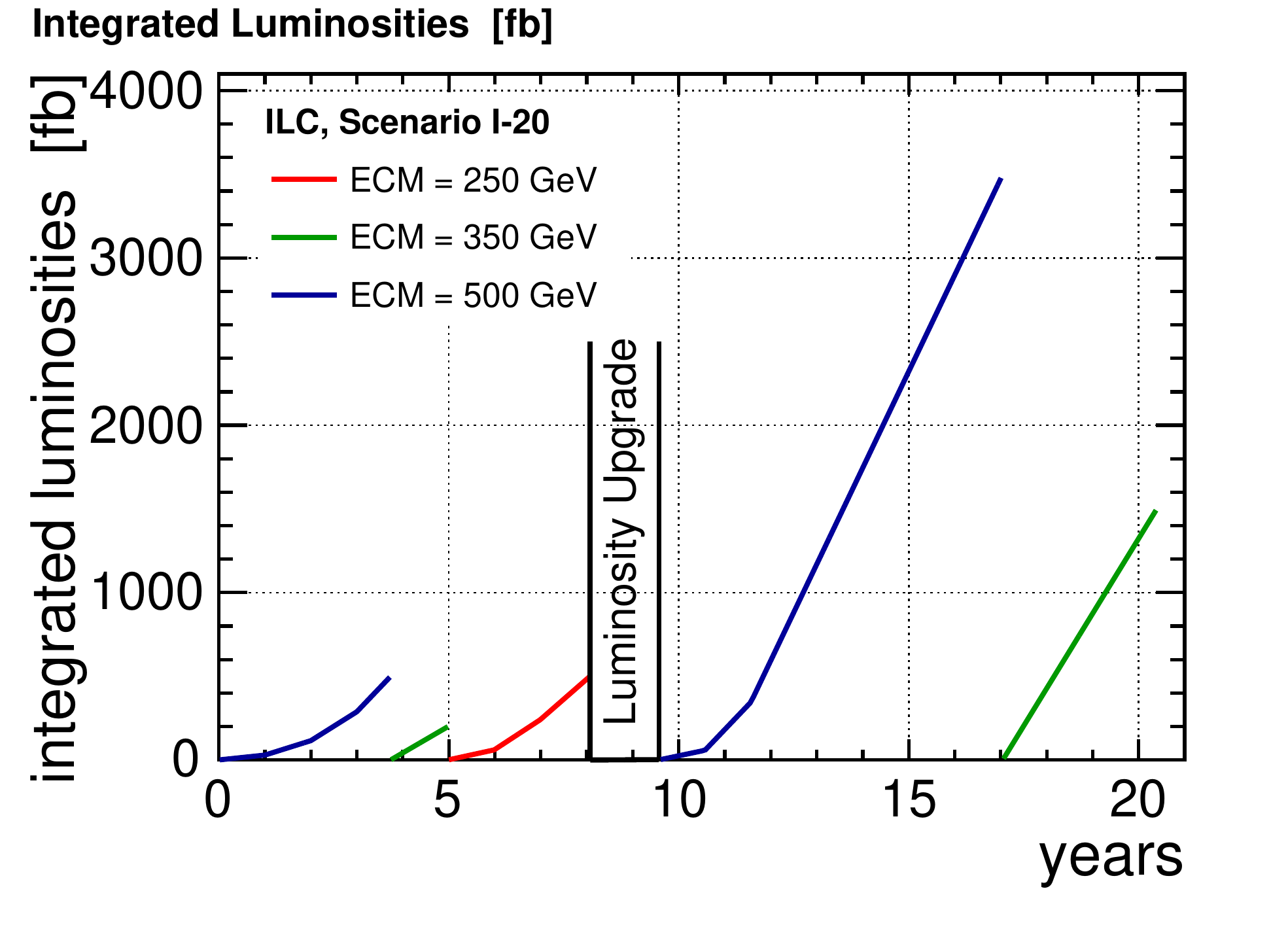}
  \caption{Accumulation of integrated luminosity versus real time for scenario I-20.
  \label{fig:realtime-I-20}}
\end{figure}

\begin{figure}[b]
\centering
\includegraphics[width=0.85\textwidth]{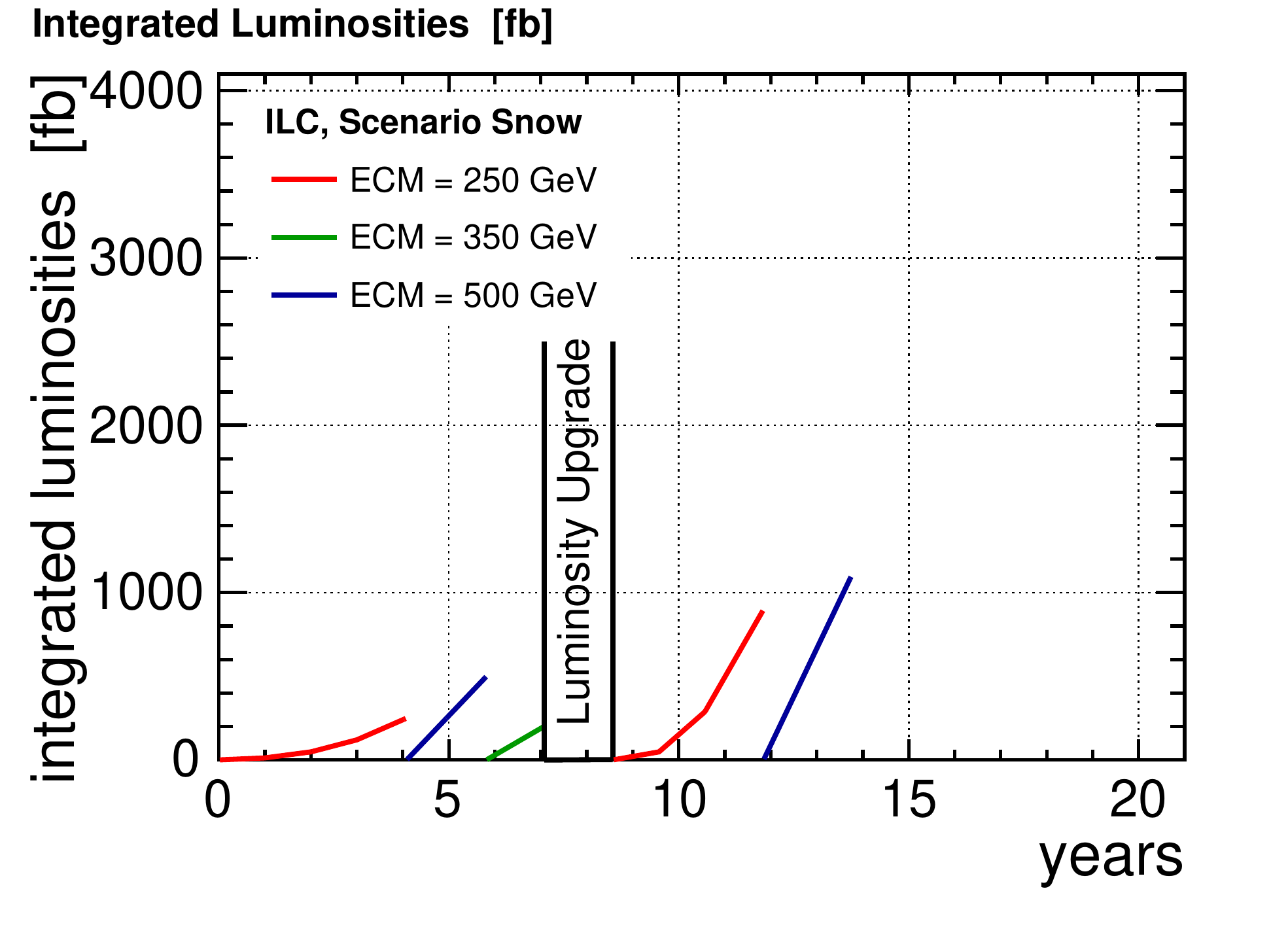}
  \caption{Accumulation of integrated luminosity versus real time for scenario Snow.
  \label{fig:realtime-Snow}}
\end{figure}

{\color{myblue}
The timelines for integrated luminosity for each of the operating scenarios 
(G-20, H-20 , I-20 and Snow) are represented in Figures~\ref{fig:realtime-G-20} through~\ref{fig:realtime-Snow}. Each figure provides a
graph of integrated luminosity versus calendar time at the respective centre-of-mass energies (colour coded). Tables~\ref{tab:realtimeG20} through~\ref{tab:realtimeSnow} show the detailed
definition of each of the scenario timelines. 

These timelines have been derived  under the following assumptions:
}

\paragraph{Basic assumptions}

\begin{itemize} 
\item All plots are presented in calendar years.
\item A full calendar year is assumed to represent eight months running at an efficiency of 75\% (the
RDR assumption). This corresponds approximately to $Y= 1.6 \times 10^7$ seconds of integrated
running. (This is significantly higher than a Snowmass year of $10^7$ seconds.)
\item t = 0 (start of Year 1) is the start of running for physics. Year 0 (-1 $\le$ t $<$ 0), directly after
construction, is assumed to be for machine commissioning only (not shown in the plots).
\item If the peak instantaneous luminosity is L, then the nominal integrated luminosity for a fully-operational
calendar year is $\int \mathcal{L} dt = L \times Y$. For any given calendar year during a period of ramp-up,
the integrated luminosity for that year is $f \times \int \mathcal{L} dt$, where $f$ is the ramp fraction associated with that year ($f \le 1$).
\end{itemize}

\paragraph{Peak luminosity assumptions}
\paragraph{}

The peak luminosities used for each centre-of-mass energy are based on the numbers published in the ILC TDR. 
However, the published
figures all assume 5 Hz collision rates. In the following scenarios, advantage has been taken of the reduced linac 
electrical power  and cryogenic loads at low-gradient operation to allow 10-Hz and 7-Hz running at 250~GeV 
and 350~GeV centre-of-mass running respectively.  For the main linac higher-rate collisions are feasible since all 
linac RF hardware has been designed for a maximum 10-Hz operation. For the baseline luminosity beam 
currents (1312 bunches per pulse), the sources and damping rings are also capable of 10-Hz operation. 
However, for the luminosity upgrade (2625 bunches per pulse), installation of additional damping ring RF beyond what has 
been 
specified in the TDR would be required. All other sub-systems should be able to 
cope with the increase in repetition rate and high beam currents, but this remains to be confirmed. 
It should be noted that 10-Hz 
collisions at 250 GeV centre-of-mass energy excludes the 10-Hz positron production mode as described in the TDR, and a longer 
positron-production undulator would be required, for which tunnel space is already foreseen in the baseline design.

\paragraph{Ramp-up assumptions}

\begin{itemize}
\item A ramp-up of luminosity performance is in general assumed after: (a) initial construction and after
`year 0' commissioning; (b) after a downtime for a luminosity upgrade; (c) a change in operational mode which may require some learning curve (e.g. going
to 10-Hz collisions).
\item A ramp is defined as a set of ramp factors $f$, one factor for each consecutive integral calendar
year at the beginning of a specific run.
\item For the initial physics run after construction and year 0 commissioning, the RDR ramp of 10\%,
30\%, 60\% and 100\% over the first four calendar years is always assumed (all scenarios).
\item  The ramp after the shutdown for installation of the luminosity upgrade is assumed slightly shorter (10\%, 50\%, 100\%) with no year 0.
\item Going down in centre of mass energy from $500$\,GeV to $350$\,GeV or $250$\,GeV is assumed to have no ramp
associated with it, since there is no modification (shutdown) to the machine. 
\item Going to 10-Hz operation at 50\% gradient does assume a ramp however (25\%, 75\%, 100\%),
since 10-Hz affects the entire machine including the damping rings and sources etc. 

\end{itemize}

\begin{table}[h]
\centering
\begin{footnotesize}
  \renewcommand{\arraystretch}{1.10}
\begin{tabular}{|l|c|c|c|c|c|c|c|c|c|l|}
\hline
      &  $\sqrt{s}$ &  $\int{\mathcal{L} dt}$ & $L_{\mathrm{peak}}$ &  \multicolumn{4}{l|}{Ramp} & $T$ & $T_{\mathrm{tot}}$ & Comment \\
\hline
            &[GeV] & [fb$^{-1}$] & [fb$^{-1}/$a]  &  1   &  2   &  3  &  4  & [a]   &   [a]      &          \\
\hline
Physics run & 500 & 1000 & 288 & 0.1  & 0.3  & 0.6 & 1.0 & 5.5 &   5.5 & TDR nominal at $5$\,Hz\\ 
Physics run & 350 &  200 & 160 & 1.0  & 1.0  & 1.0 & 1.0 & 1.2 &   6.7 & TDR nominal at $5$\,Hz\\ 
Physics run & 250 &  500 & 240 & 0.25 & 0.75 & 1.0 & 1.0 & 3.1 &   9.8 & operation at $10$\,Hz\\ 
Shutdown    &     &      &     &      &      &     &     & 1.5 &  11.3 & Luminosity upgrade\\ 
Physics run & 500 & 4000 & 576 & 0.1  & 0.5  & 1.0 & 1.0 & 8.4 &  19.7 & TDR lumi-up at $5$\,Hz\\ 
\hline
\end{tabular}
\caption{Scenario G-20: Sequence of energy stages and their real-time conditions.\label{tab:realtimeG20}} 
\end{footnotesize}
\end{table}

\begin{table}[h]
\centering
\begin{footnotesize}
  \renewcommand{\arraystretch}{1.10}
\begin{tabular}{|l|c|c|c|c|c|c|c|c|c|l|}
\hline
      &  $\sqrt{s}$ &  $\int{\mathcal{L} dt}$ & $L_{\mathrm{peak}}$ &  \multicolumn{4}{l|}{Ramp} & $T$ & $T_{\mathrm{tot}}$ & Comment \\
\hline
            &[GeV] & [fb$^{-1}$] & [fb$^{-1}/$a]  &  1   &  2   &  3  &  4  & [a]   &   [a]      &          \\
\hline
Physics run & 500 &  500 & 288 & 0.1  & 0.3  & 0.6 & 1.0 & 3.7  &  3.7 & TDR nominal at $5$\,Hz\\ 
Physics run & 350 &  200 & 160 & 1.0  & 1.0  & 1.0 & 1.0 & 1.3  &  5.0 & TDR nominal at $5$\,Hz\\ 
Physics run & 250 &  500 & 240 & 0.25 & 0.75 & 1.0 & 1.0 & 3.1  &  8.1 & operation at $10$\,Hz\\ 
Shutdown    &     &      &     &      &      &     &     & 1.5  &  9.6 & Luminosity upgrade\\ 
Physics run & 500 & 3500 & 576 & 0.1  & 0.5  & 1.0 & 1.0 & 7.4  & 17.0 & TDR lumi-up at $5$\,Hz\\ 
Physics run & 250 & 1500 & 480 & 1.0  & 1.0  & 1.0 & 1.0 & 3.2  & 20.2 & lumi-up operation at $10$\,Hz\\ 
\hline
\end{tabular}
\caption{Scenario H-20: Sequence of energy stages and their real-time conditions.\label{tab:realtimeH20}} 
\end{footnotesize}
\end{table}

\begin{table}[h]
\centering
\begin{footnotesize}
  \renewcommand{\arraystretch}{1.10}
\begin{tabular}{|l|c|c|c|c|c|c|c|c|c|l|}
\hline
      &  $\sqrt{s}$ &  $\int{\mathcal{L} dt}$ & $L_{\mathrm{peak}}$ &  \multicolumn{4}{l|}{Ramp} & $T$ & $T_{\mathrm{tot}}$ & Comment \\
\hline
            &[GeV] & [fb$^{-1}$] & [fb$^{-1}/$a]  &  1   &  2   &  3  &  4  & [a]   &   [a]      &          \\
\hline
Physics run & 500 &  500 & 288 & 0.1  & 0.3  & 0.6 & 1.0 & 3.7 &  3.7  & TDR nominal at $5$\,Hz\\ 
Physics run & 350 &  200 & 160 & 1.0  & 1.0  & 1.0 & 1.0 & 1.3 &  5.0  & TDR nominal at $5$\,Hz\\ 
Physics run & 250 &  500 & 240 & 0.25 & 0.75 & 1.0 & 1.0 & 3.1 &  8.1  & operation at $10$\,Hz\\ 
Shutdown    &     &      &     &      &      &     &     & 1.5 &  9.6  & Luminosity upgrade\\ 
Physics run & 500 & 3500 & 576 & 0.1  & 0.5  & 1.0 & 1.0 & 7.4 & 17.0  & TDR lumi-up at $5$\,Hz\\ 
Physics run & 350 & 1500 & 448 & 1.0  & 1.0  & 1.0 & 1.0 & 3.4 & 20.4  & lumi-up operation at $7$\,Hz\\ 
\hline
\end{tabular}
\caption{Scenario I-20: Sequence of energy stages and their real-time conditions.\label{tab:realtimeI20}} 
\end{footnotesize}
\end{table}

\begin{table}[h]
\centering
\begin{footnotesize}
  \renewcommand{\arraystretch}{1.10}
\begin{tabular}{|l|c|c|c|c|c|c|c|c|c|l|}
\hline
      &  $\sqrt{s}$ &  $\int{\mathcal{L} dt}$ & $L_{\mathrm{peak}}$ &  \multicolumn{4}{l|}{Ramp} & $T$ & $T_{\mathrm{tot}}$ & Comment \\
\hline
            &[GeV] & [fb$^{-1}$] & [fb$^{-1}/$a]  &  1   &  2   &  3  &  4  & [a]   &   [a]      &          \\
\hline
Physics run & 250 &  250 & 120 & 0.1  & 0.3  & 0.6 & 1.0 & 4.1   &  4.1  & TDR nominal at $5$\,Hz\\ 
Physics run & 500 &  500 & 288 & 1.0  & 1.0  & 1.0 & 1.0 & 1.7   &  5.8  & TDR nominal at $5$\,Hz\\ 
Physics run & 350 &  200 & 160 & 1.0  & 1.0  & 1.0 & 1.0 & 1.3   &  7.1  & TDR nominal at $5$\,Hz\\ 
Shutdown    &     &      &     &      &      &     &     & 1.5   &  8.6  & Luminosity upgrade\\ 
Physics run & 250 &  900 & 480 & 0.1  & 0.5  & 1.0 & 1.0 & 3.0   & 11.8  & lumi-up operation at $10$\,Hz\\ 
Physics run & 500 & 1100 & 576 & 1.0  & 1.0  & 1.0 & 1.0 & 2.0   & 13.8  & TDR lumi-up at $5$\,Hz\\ 
\hline
\end{tabular}
\caption{Scenario Snow: Sequence of energy stages and their real-time conditions.\label{tab:realtimeSnow}} 
\end{footnotesize}
\end{table}

\paragraph{Shutdowns}

\begin{itemize}

\item A major 18 month shutdown is assumed for the luminosity upgrade.
\item The shutdown is for the TDR luminosity upgrade, where the number of bunches per pulse
is increased from 1312 to 2625. This requires the installation of an additional 50\% of klystrons
and modulators, as well as the possible installation of a second positron damping ring. It is
assumed that linac and damping ring installation occur in parallel and do not interfere with each
other.
\item This down-time may be on the optimistic side, but would appear to be roughly
consistent with the TDR construction installation rates, assuming that the same level of
manpower is available, and that all the necessary components for installation are (mostly)
available at the time the shutdown starts.

\end{itemize}

%

%% file: physics.tex
\section{Time Development of Physics Results}
\label{sec:physics}

In this section we present some examples of how important physics
results evolve in time for the three 
scenarios presented above.
All plots in this section are preliminary since not all analyses involved have been finished yet,
so that some measurements are extrapolated, e.g. from other center-of-mass energies.

\subsection{Higgs couplings to fermions and gauge bosons}
\label{subsec:physics_higgs}

\begin{figure}[htbp]
\centering
\includegraphics[width=0.9\textwidth]{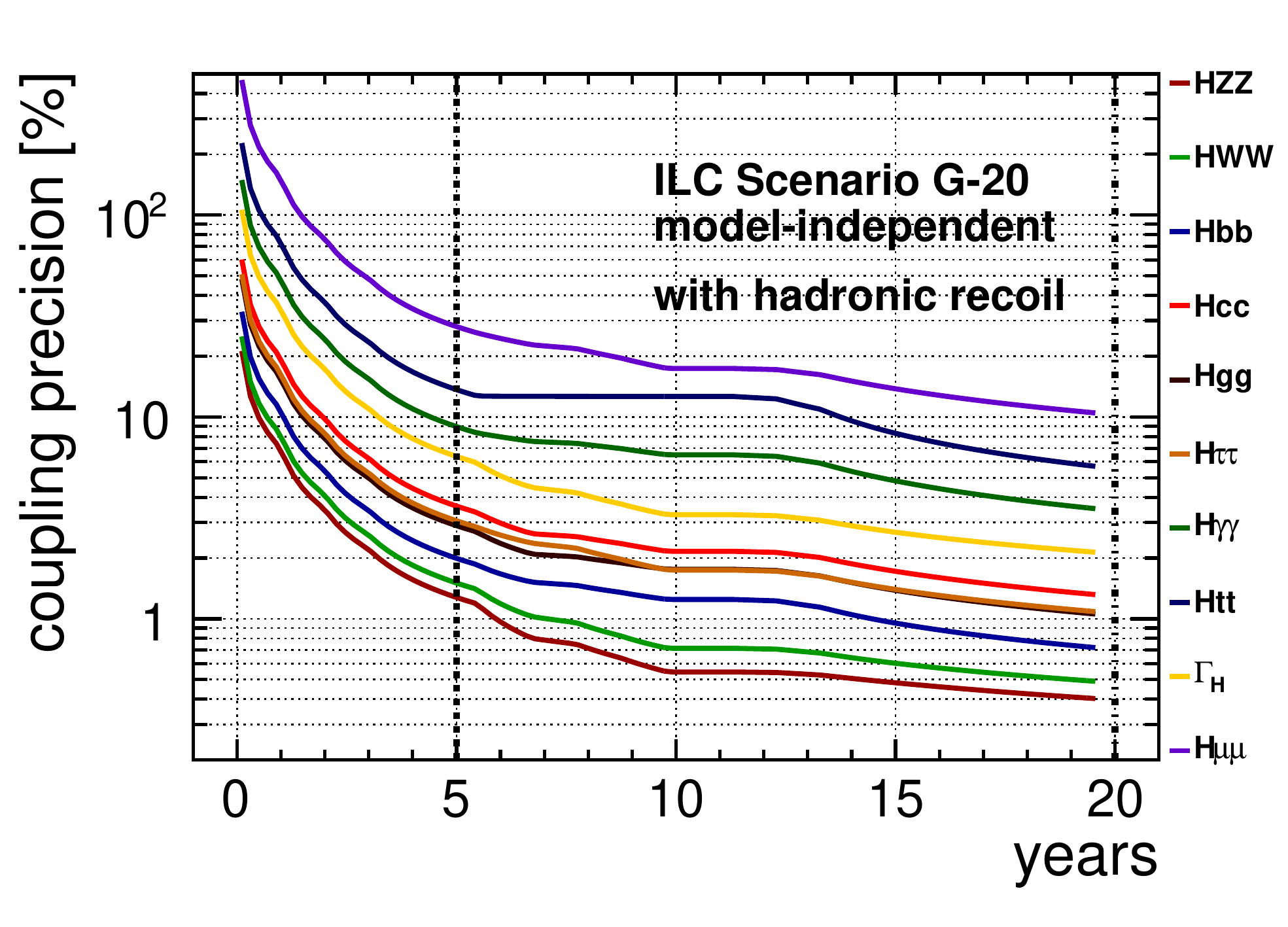}
\includegraphics[width=0.9\textwidth]{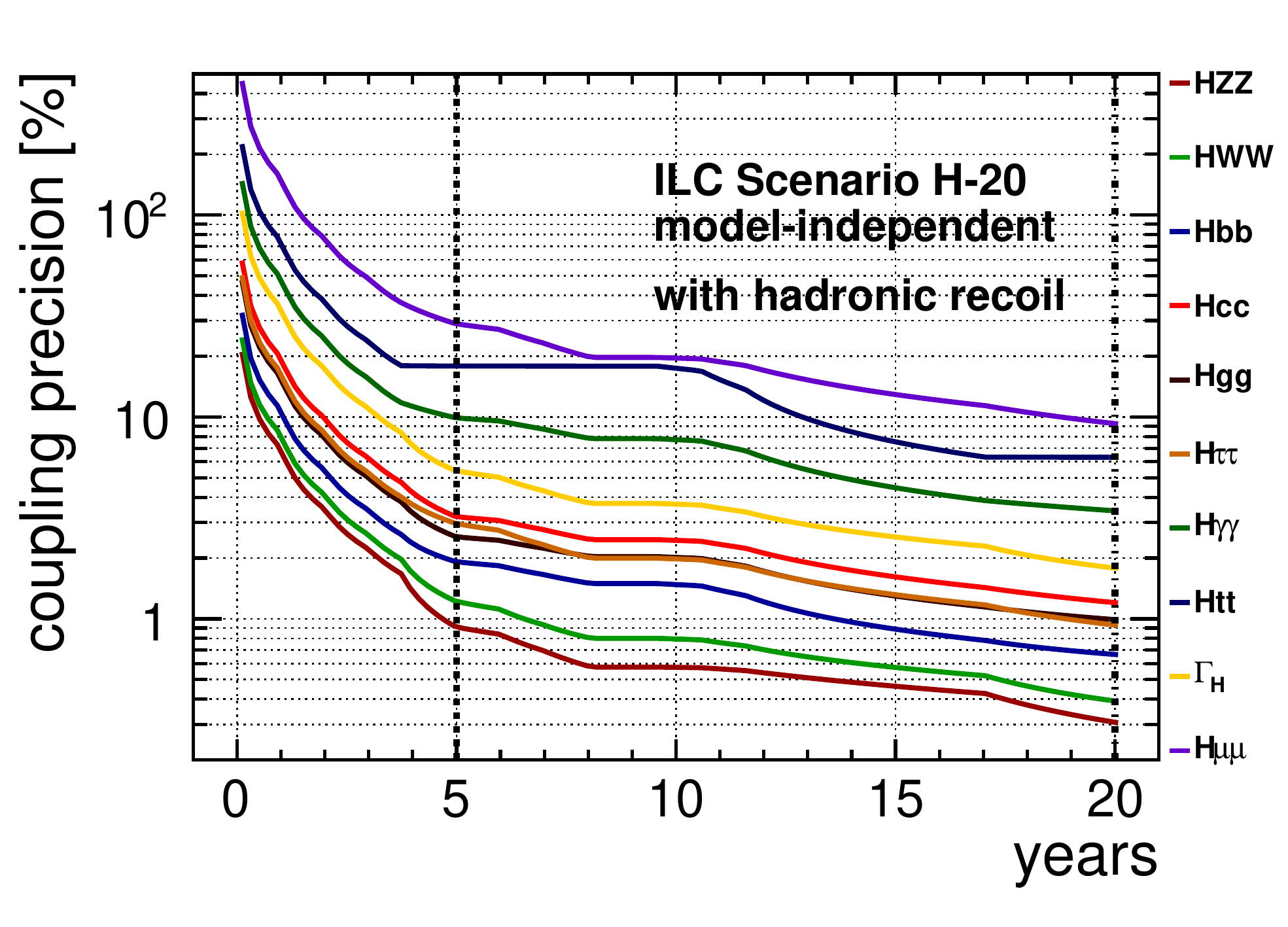}
  \caption{\color{myblue} Time evolution of precision on various couplings of the Higgs boson in
  the scenarios G-20 and H-20. \label{fig:HiggsCouplingsGH}}
\end{figure}

\begin{figure}[htbp]
\centering
\includegraphics[width=0.9\textwidth]{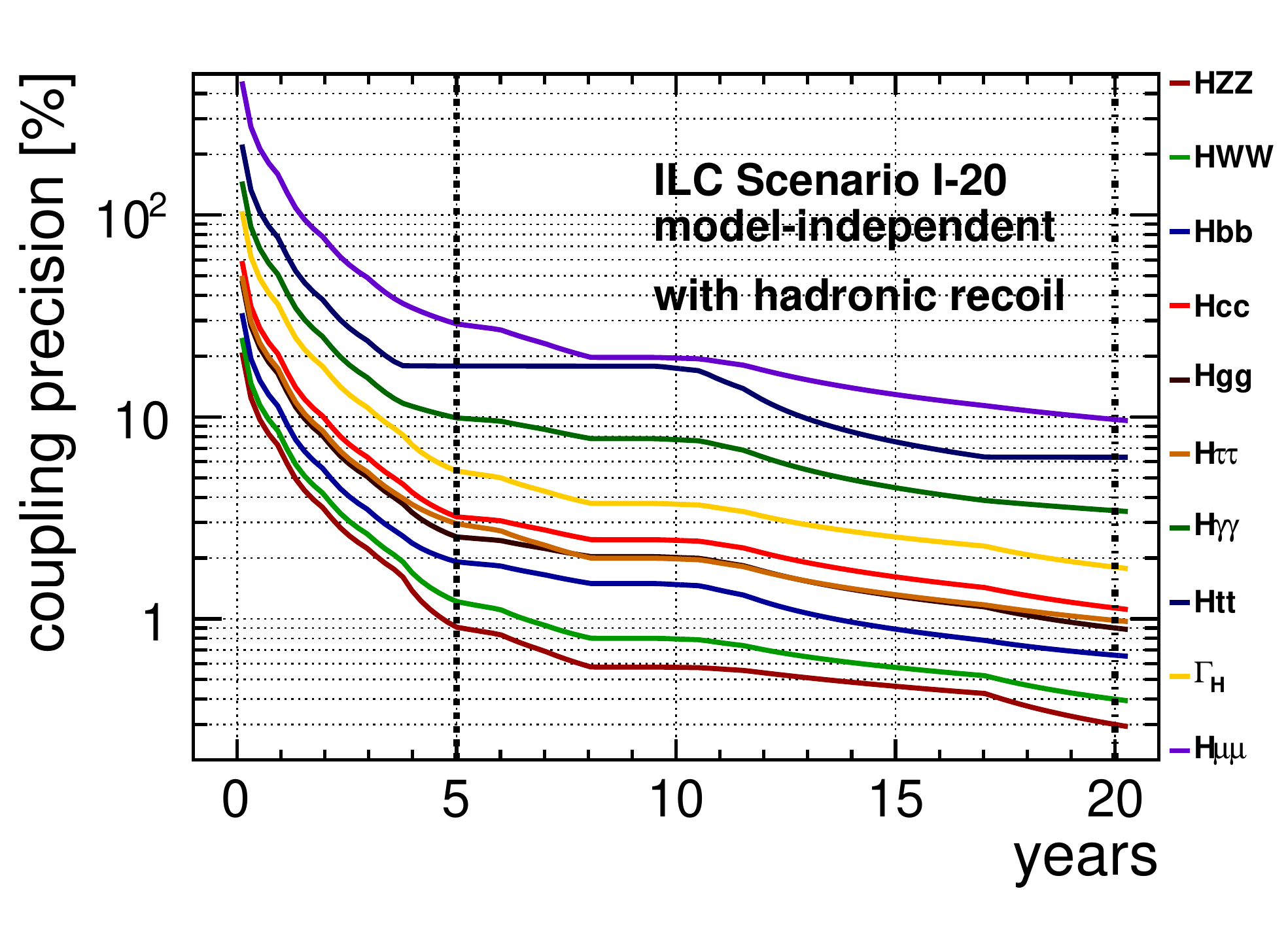}
\includegraphics[width=0.9\textwidth]{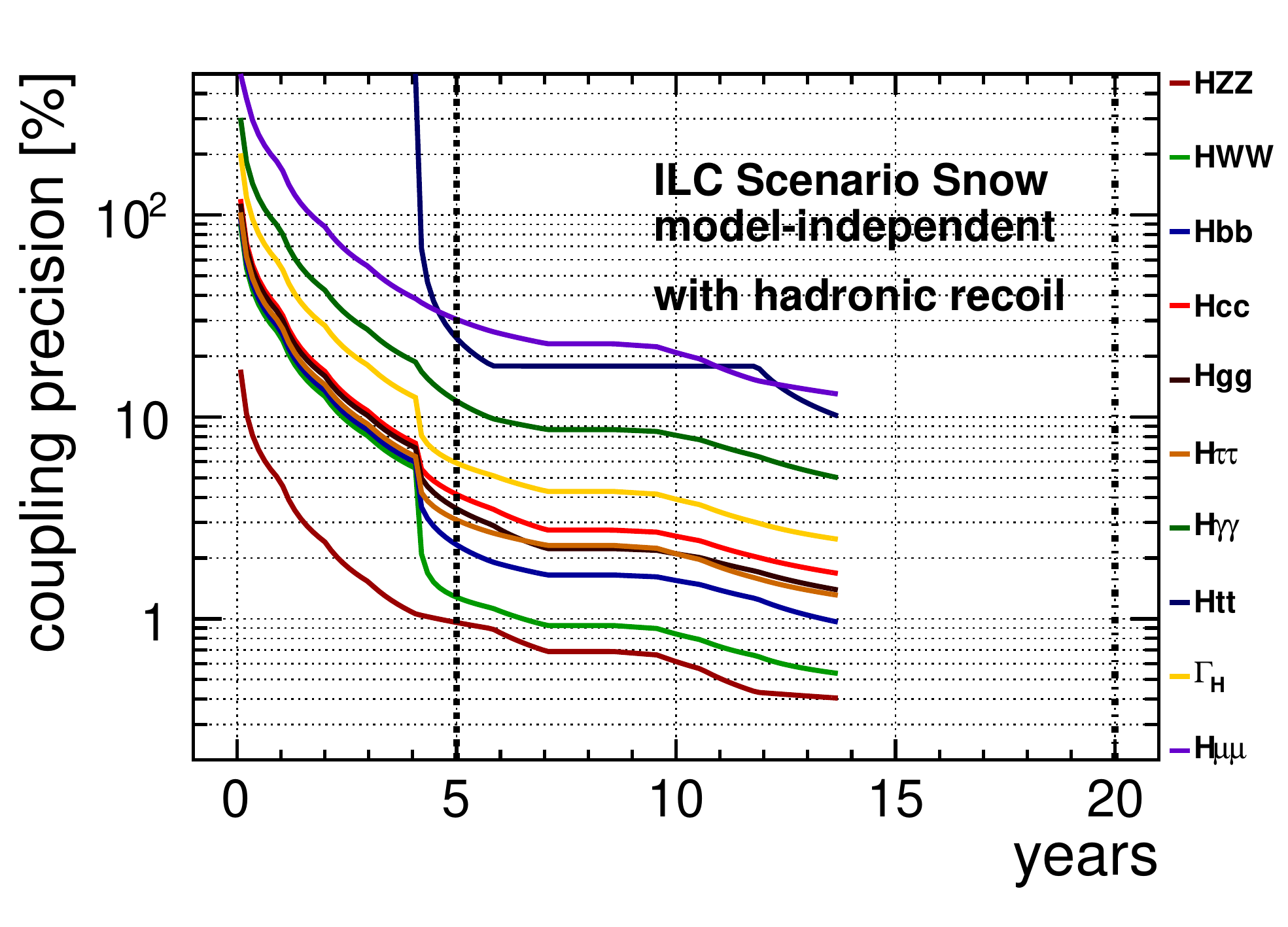}
  \caption{\color{myblue} Time evolution of precision on various couplings of the Higgs boson in the I-20 and Snow scenarios. \label{fig:HiggsCouplingsISnow}}
\end{figure}

{\color{myblue}
Figures~\ref{fig:HiggsCouplingsGH} and~\ref{fig:HiggsCouplingsISnow} show the current snapshot of available Higgs studies interpreted 
in a fully model-independent global fit. It is in many cases based
on the same full-simulation studies used in the Snowmass ILC Higgs Whitepaper~\cite{Asner:2013psa}. However at the time of Snowmass,
studies at $\sqrt{s}=350$\,GeV were not yet available, as well as
results for $m_h = 125$\,GeV and other beam polarisations then $P(e^-,e^+)=(-80\%,+30\%)$.
These have been added meanwhile and their preliminary results are included in our fits. 
In appendix~\ref{sec:higgsmeas}, we give the full set of $\sigma \times BR$ projections
which we use as input to the coupling fit.}


Another development since Snowmass are
recent studies~\cite{Thomson2014:hadronrecoil} 
which showed that a measurement of the Higgstrahlung cross section using the events with a Higgs recoiling from the 
hadronic decay of the $Z$ boson  
(referred to as the hadronic recoil measurement) 
can be performed in a nearly model-independent way at $\sqrt{s}=350$\,GeV,
in the sense that detection efficiencies for SM Higgs decays differ by no more than 7\%. This translates
into a systematic error for the  model-dependency of less than 11\% of the statistical uncertainty\cite{Barklow:2014hadsys}. Therefore we generally 
include the hadronic recoil measurements in the coupling fits.

Another method is to use the constraint  $\sigma(ZH)=\sum_{i} {\sigma(ZH)\cdot BR_i}$, where the sum is over all Higgs decays, including Beyond the Standard Model
(BSM) decays.  This constraint is model independent if 
the measurement error for the branching ratio of BSM Higgs
decays is included in the fit.  If in the future an analysis can deliver a precision of 
$BR(H\rightarrow BSM) < 0.9\%$ at 95\% C.L., then this constraint can lead to 
further improvements in the Higgs coupling precision~\cite{Peskin:2013xra}.  


{\color{myblue}
The scenarios G-20, H-20 and I-20 show rather similar performance. At a closer look, G-20 performs slighty
worse for most couplings due to the later luminosity upgrade and thus somewhat lower integrated luminosities.
Scenarios H-20 and I-20, which differ only in a final longer run at $\sqrt{s}=250$\,GeV vs $350$\,GeV,
look almost identical. However since at the time of writing this document it has not yet been demonstrated that
the Higgs mass can be measured with sufficient precision from kinematic reconstruction (c.f.\ \ref{subsec:physics_Mhiggs}), we conservatively consider 
H-20 the prefered scenario.  

The scenario ``Snow'' differs most from the other scenarios due to its initial run at $\sqrt{s}=250$\,GeV,
which is beneficial for the precision on the $HZZ$ coupling. Most other coupling precisions are limited by 
the knowledge of the $HWW$ coupling or by statistics, and thus only reach their full potential once $\sqrt{s}\ge350$\,GeV. Obviously the direct measurement of the top Yukawa coupling is only possible once
$\sqrt{s}\ge500$\,GeV.
Thus the early physics output will be limited when starting operation at $\sqrt{s}=250$\,GeV.
}

\begin{figure}[htbp]
\centering
\includegraphics[width=0.9\textwidth]{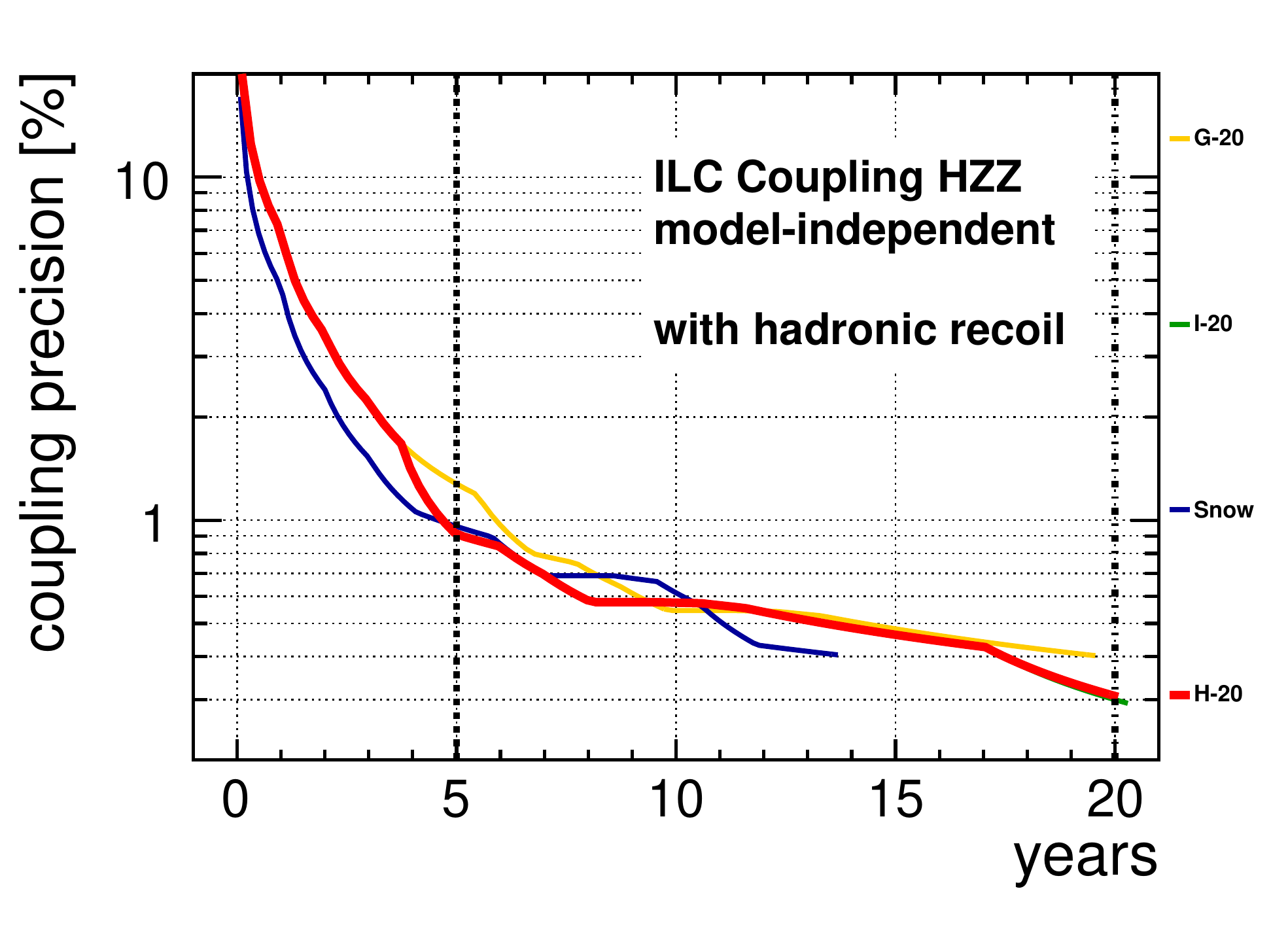}
\includegraphics[width=0.9\textwidth]{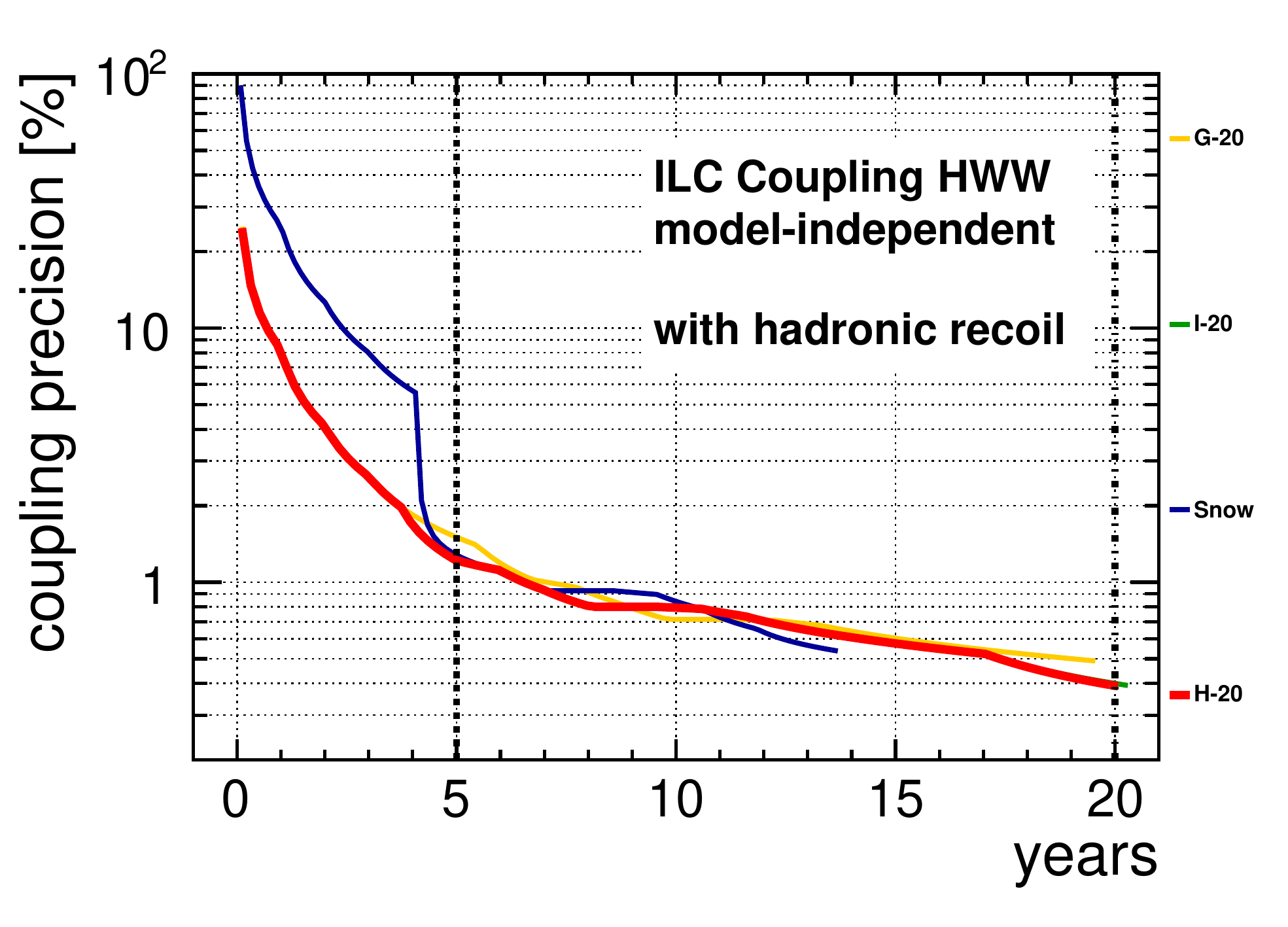}
  \caption{\color{myblue} Time evolution of precision on $g_{HZZ}$ and $g_{HWW}$ in all four
   scenarios. \label{fig:HiggsCouplingsHZZHWW}}
\end{figure}

\begin{figure}[htbp]
\centering
\includegraphics[width=0.9\textwidth]{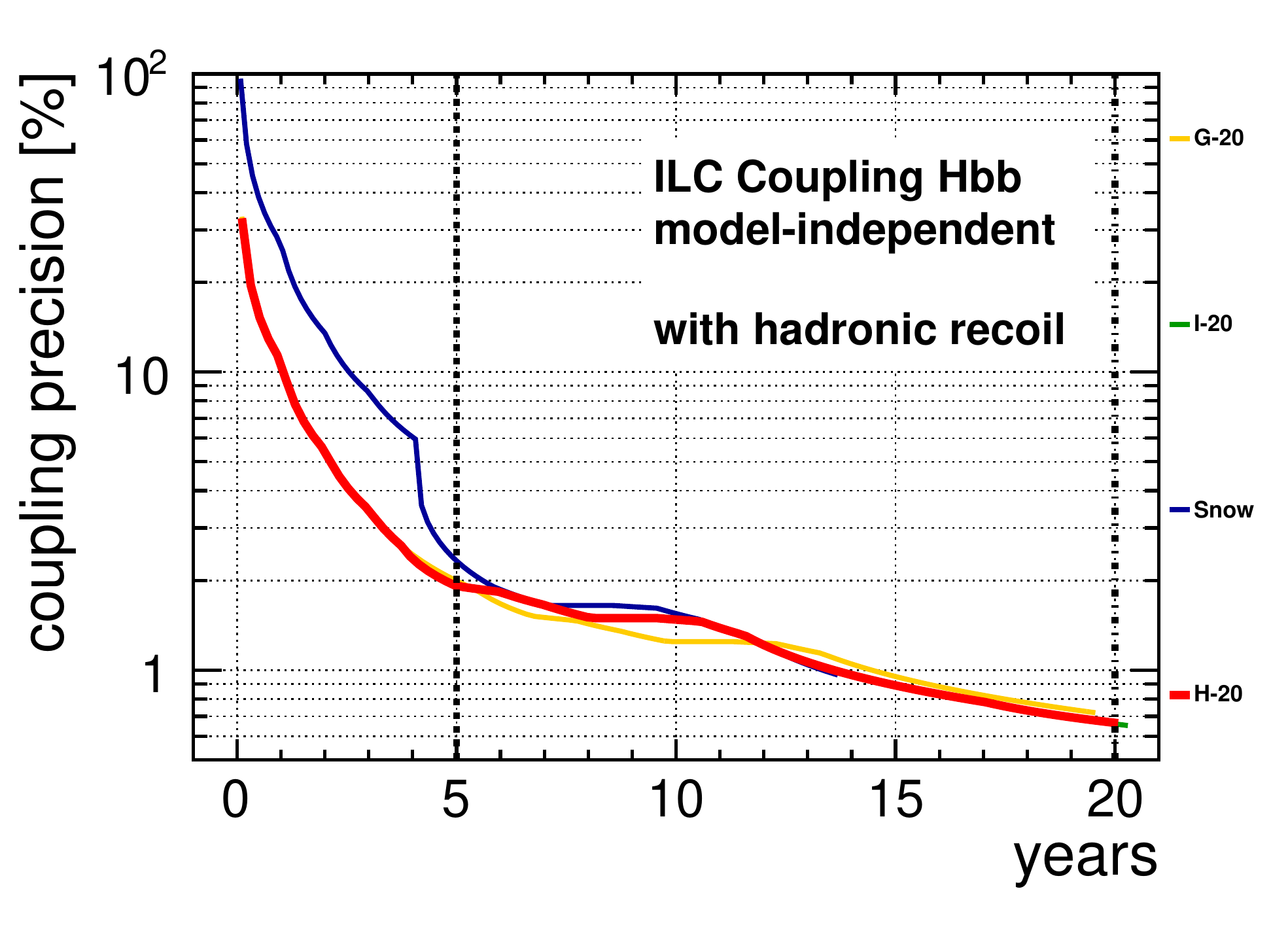}
\includegraphics[width=0.9\textwidth]{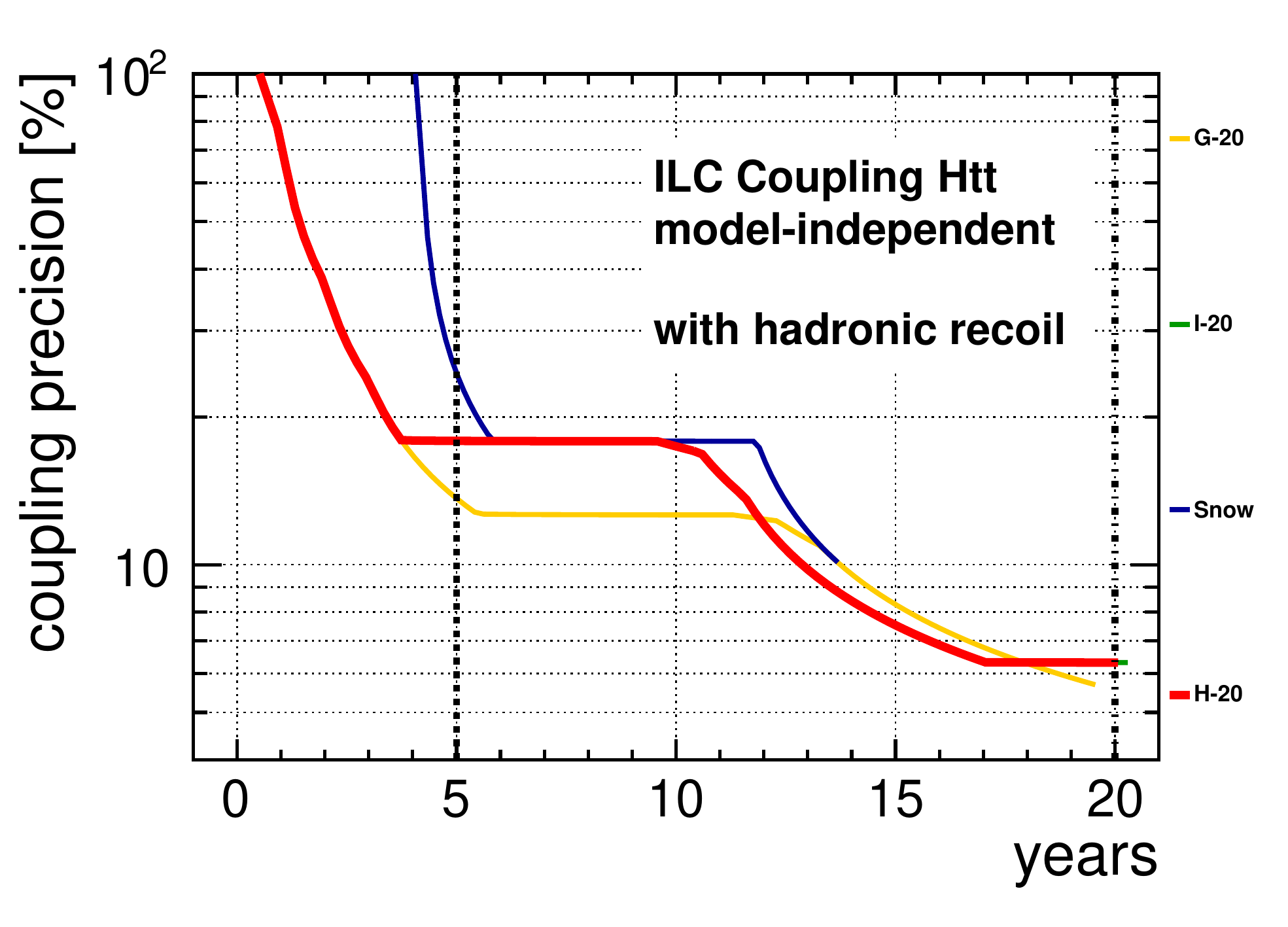}
  \caption{\color{myblue} Time evolution of precision on $g_{Hbb}$ and $g_{Htt}$ in all four
   scenarios. \label{fig:HiggsCouplingsHbbHtt}}
\end{figure}

\begin{figure}[htbp]
\centering
\includegraphics[width=0.9\textwidth]{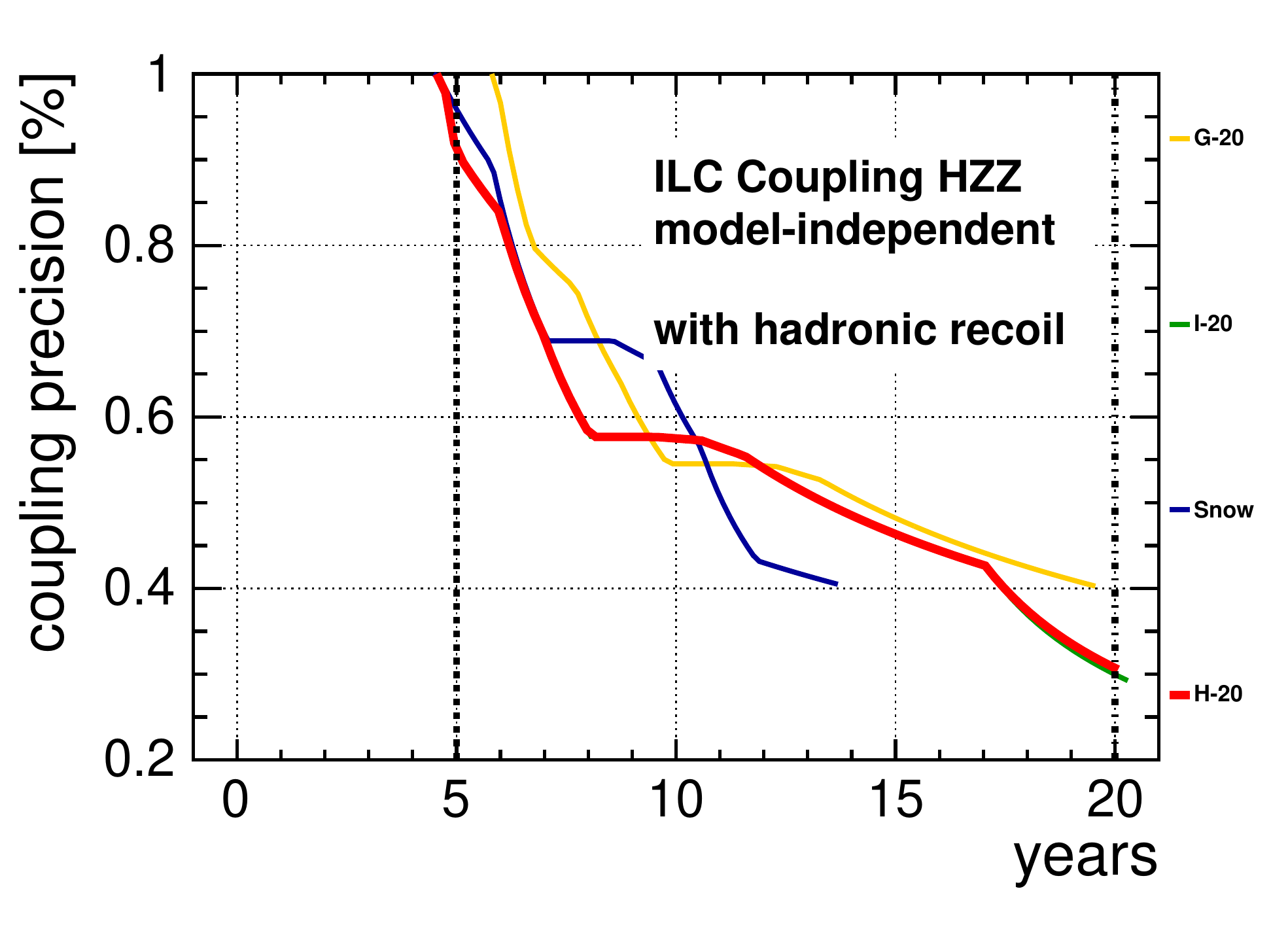}
\includegraphics[width=0.9\textwidth]{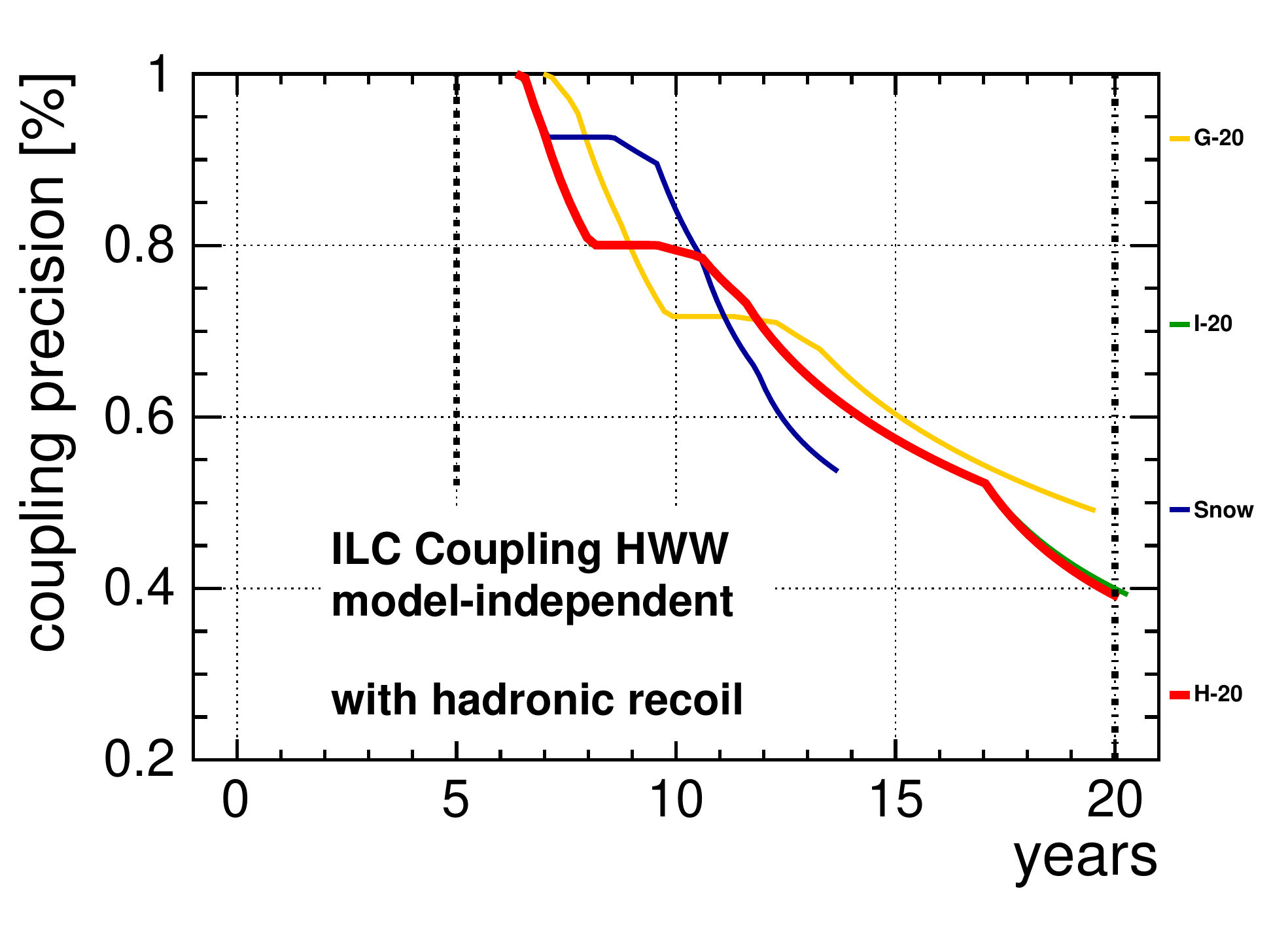}
  \caption{\color{myblue} Zoom into the time evolution of precision on $g_{HZZ}$ and $g_{HWW}$ in all four
   scenarios on a linear scale.\label{fig:HiggsCouplingsHZZHWWlin}}
\end{figure}

\begin{figure}[htbp]
\centering
\includegraphics[width=0.9\textwidth]{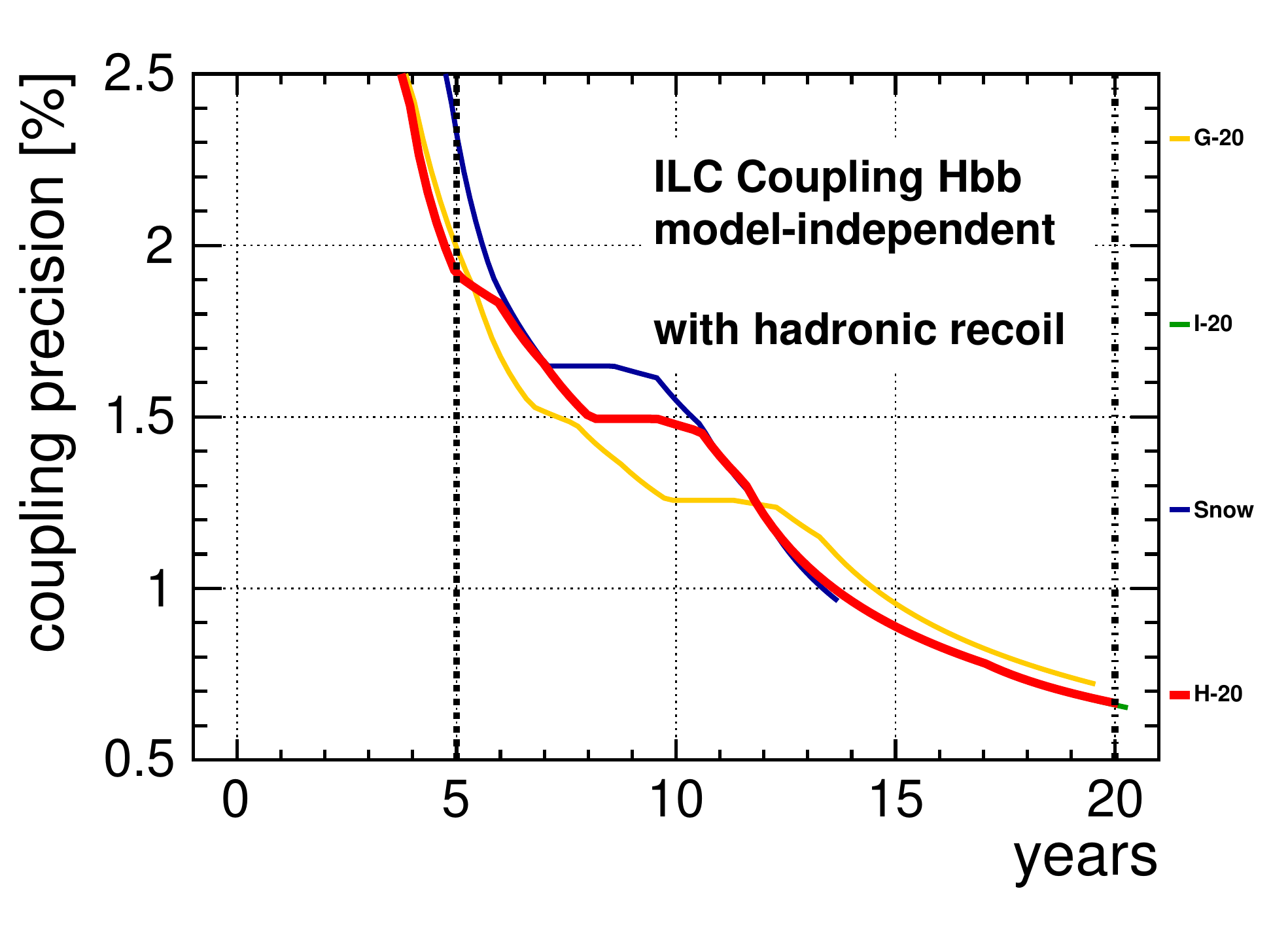}
\includegraphics[width=0.9\textwidth]{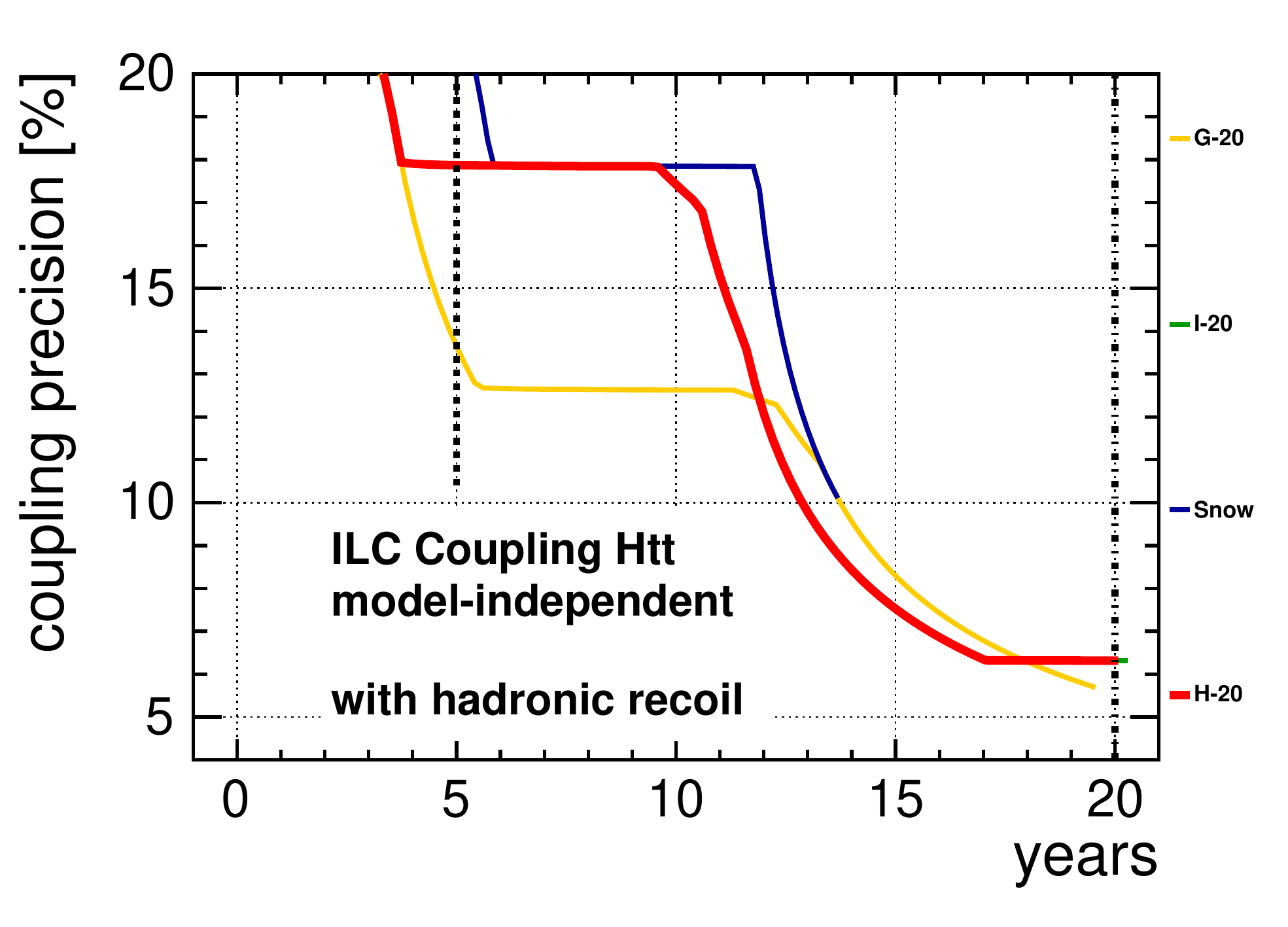}
  \caption{\color{myblue} Zoom into the time evolution of precision on $g_{Hbb}$ and $g_{Htt}$ in all four
   scenarios on a linear scale. \label{fig:HiggsCouplingsHbbHttlin}}
\end{figure}

{\color{myblue}
Figures~\ref{fig:HiggsCouplingsHZZHWW} and~\ref{fig:HiggsCouplingsHbbHtt} directly compare the performance
of the four scenarios in case of $g_{HZZ}$, $g_{HWW}$, $y_b$ and $y_t$.
Figures~\ref{fig:HiggsCouplingsHZZHWWlin} and~\ref{fig:HiggsCouplingsHbbHttlin} show the same results  
but on a linear scale in order to give a clearer picture of the precision evolution in the final years.
These figures illustrate that an early
run at lower energies as in scenario ``Snow'' is beneficial for early precision on $g_{HZZ}$, while the 
other couplings profit more from starting at $500$\,GeV. Clearly scenario G-20 gives the best precision 
of the top Yukawa coupling before the luminosity upgrade. It is interesting to note that although $g_{HWW}$ profits significantly from initial operation at $500$\,GeV (c.f.\ Fig~\ref{fig:HiggsCouplingsHZZHWW} scenario ``Snow'' vs the others),
it finally becomes limited by the knowledge on $g_{HZZ}$. Thus, also the precision on $g_{HWW}$ improves
significantly from a final high-luminosity run at $250$\,GeV (c.f.\ Fig~\ref{fig:HiggsCouplingsHZZHWWlin} scenario H-20 vs G-20). 

}

\subsection{Higgs mass}
\label{subsec:physics_Mhiggs}
\begin{figure}[htbp]
\centering
\includegraphics[width=0.6\textwidth]{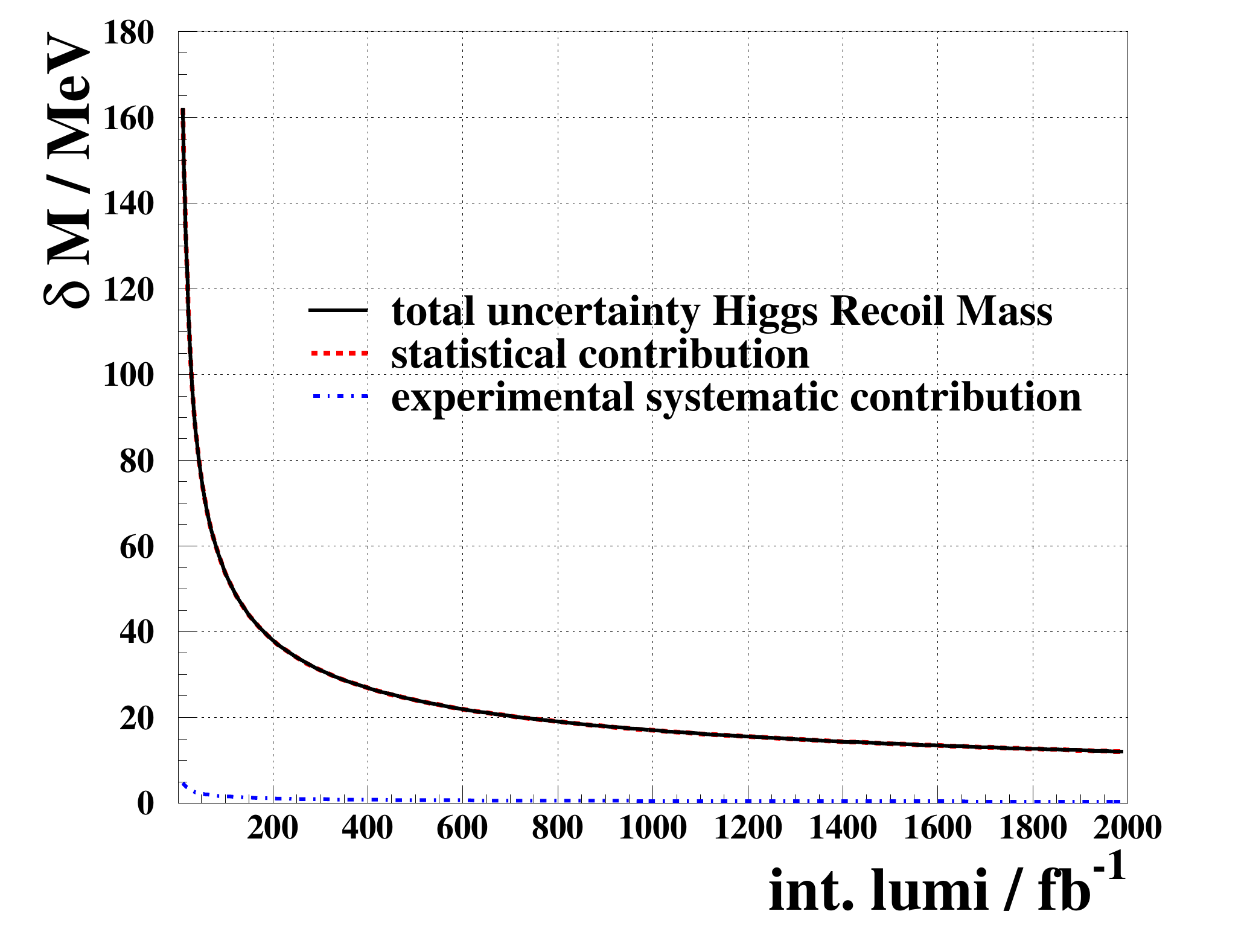}
  \caption{{Luminosity scaling of the Higgs recoil mass measurement at $\sqrt{s}$ = 250 GeV. Based on~\cite{Hengne} and~\cite{Graham}.} \label{fig:mhiggs}}
\end{figure}

Besides being an important parameter of the Standard Model in its own right, the mass of the
Higgs boson enters into the calculation of the phase space for Higgs decays,
in particular for decays into $WW$ and $ZZ$.
Thus the uncertainty on the Higgs mass translates into a parametric uncertainty on the
couplings extracted from the measurements of these decay modes. 
An uncertainty of $\delta M_H = 200$\,MeV 
(the LHC expects to reach this level of precision or better\cite{Dawson:2013bba}) 
has been estimated to cause an uncertainty of $2.2\%$ and $2.5\%$ on the partial widths of $H\to WW$ and $HH\to ZZ$, respectively~\cite{Asner:2013psa}. If one would like
to keep the parametric uncertainty on these observables at the level of $0.2\%$, $\delta M_H = 20$\,MeV would be required. Currently the only way
to reach this level of precision which has been demonstrated in full detector simulation is the Higgs recoil mass measurement with $Z \to \mu \mu$ at $\sqrt{s}=250$\,GeV. With a momentum scale calibration from $Z\to\mu\mu$ at the $Z$ pole and an in-situ beam energy calibration from $\mu\mu\gamma$ events, systematic uncertainties should be controlled at the $1$\,MeV level~\cite{Graham}. Figure~\ref{fig:mhiggs} shows the luminosity 
scaling of Higgs recoil mass uncertainty. With $500$\,fb$^{-1}$ of data
collected at $\sqrt{s}=250$\,GeV, $\delta M_H = 25$\,MeV is reached.
In addition, preliminary studies of direct reconstruction of the Higgs mass from its decays to $b\bar{b}$ and $WW$ at $\sqrt{s}=500$\,GeV show a competitive potential~\cite{Graham}. These studies should be substantiated in the future.

\subsection{Electroweak couplings of the top quark}
\begin{figure}[htb]
\centering
\includegraphics[width=0.725\textwidth]{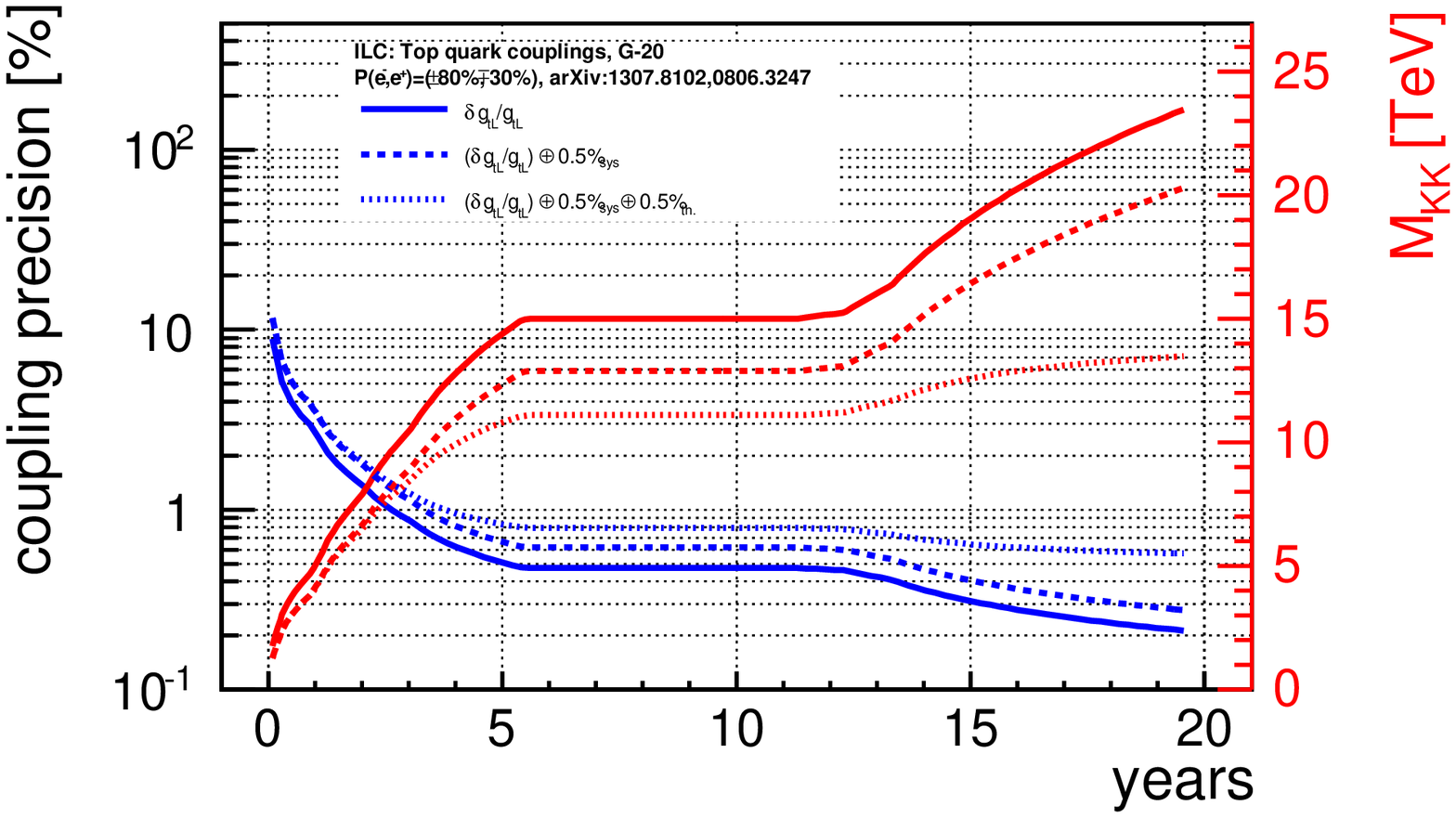}
\includegraphics[width=0.725\textwidth]{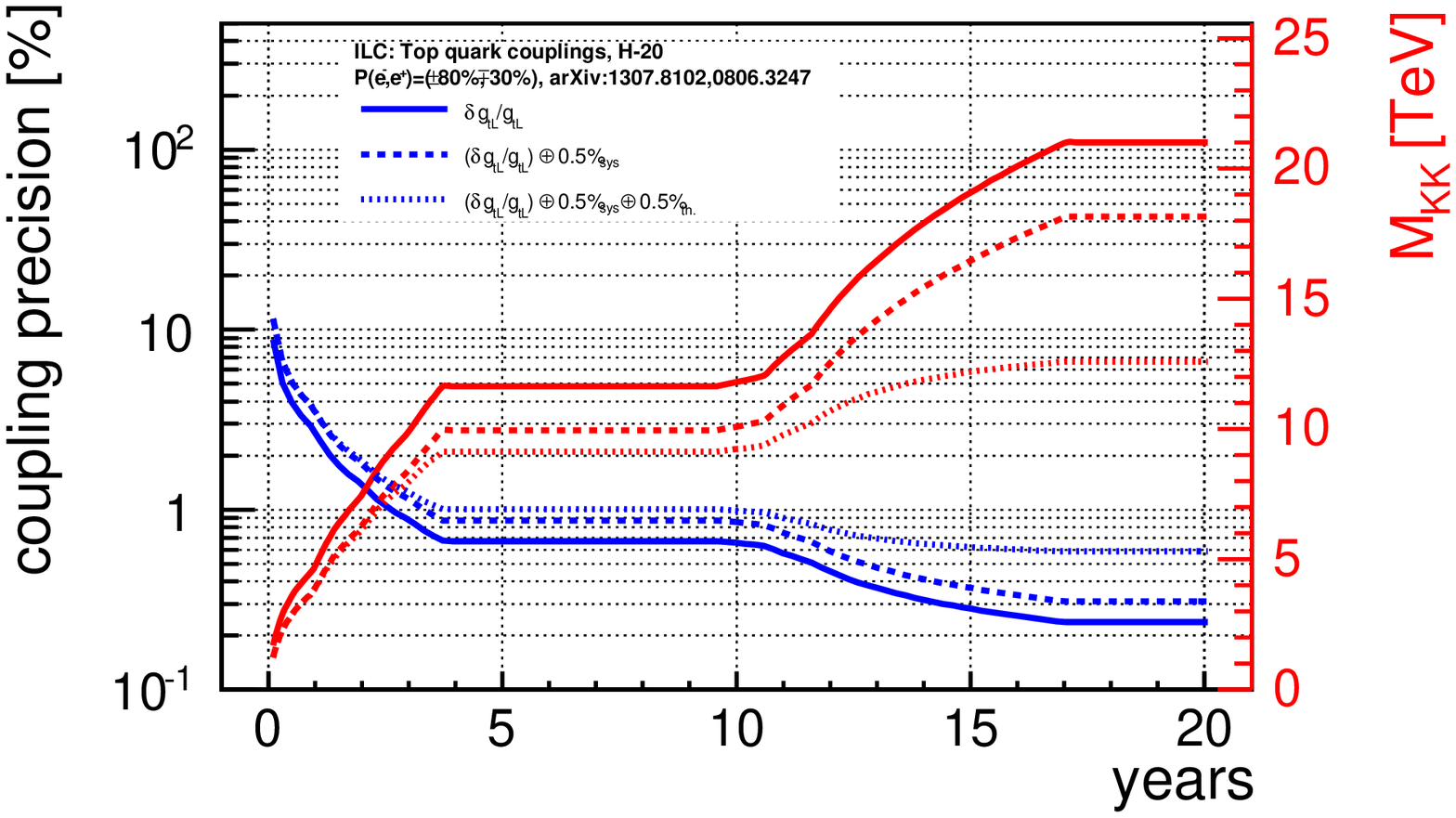}
  \caption{\color{myblue} Time-evolution of left-handed top coupling and derived from this the sensitivity to the mass scale of Kaluza-Klein excitations in an extra-dimension model~\cite{Richard:2014upa} for scenarios G-20 and H-20.\label{fig:topcoup_GH}}
\end{figure}

\begin{figure}[htb]
\centering
\includegraphics[width=0.725\textwidth]{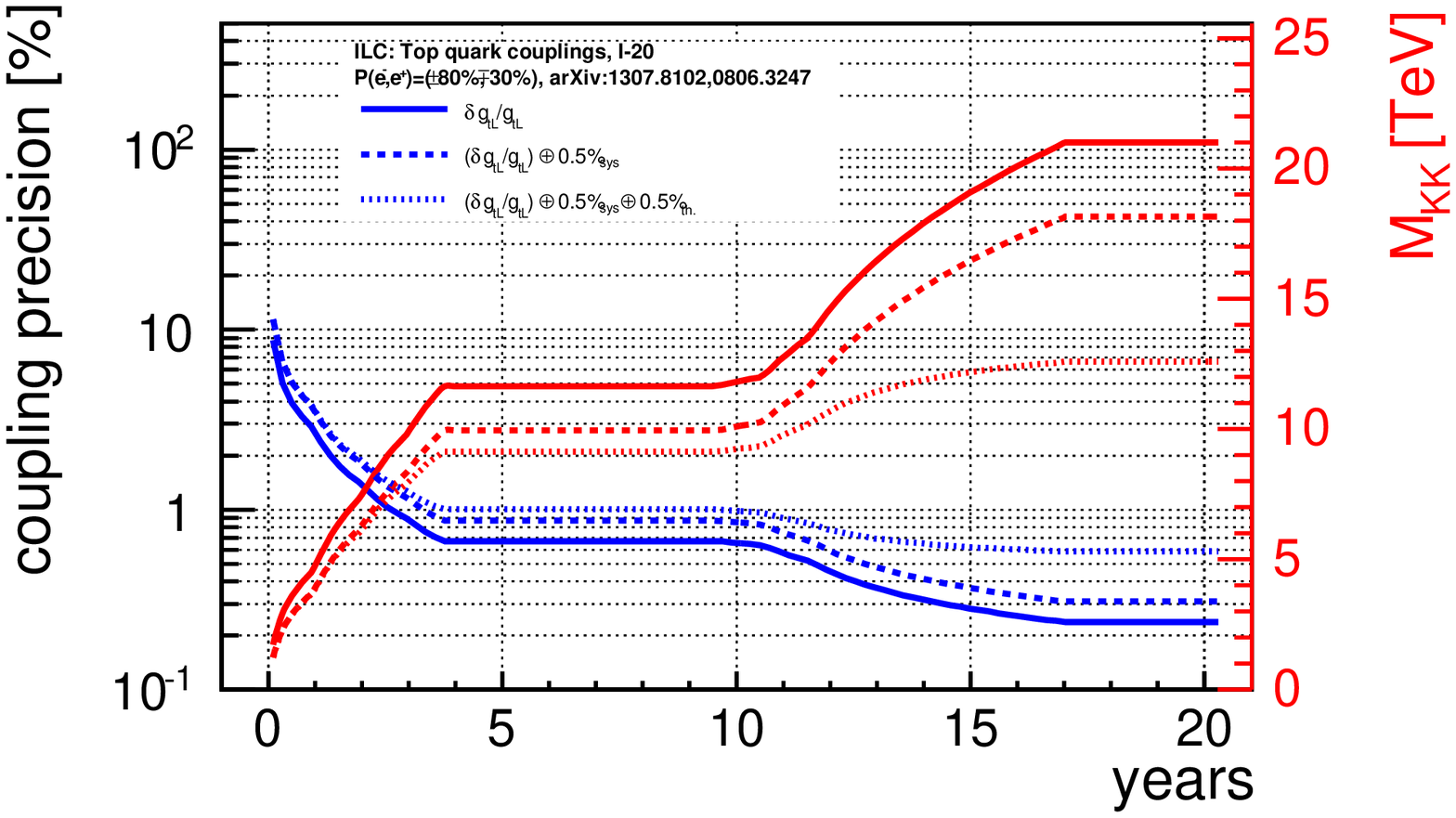}
\includegraphics[width=0.725\textwidth]{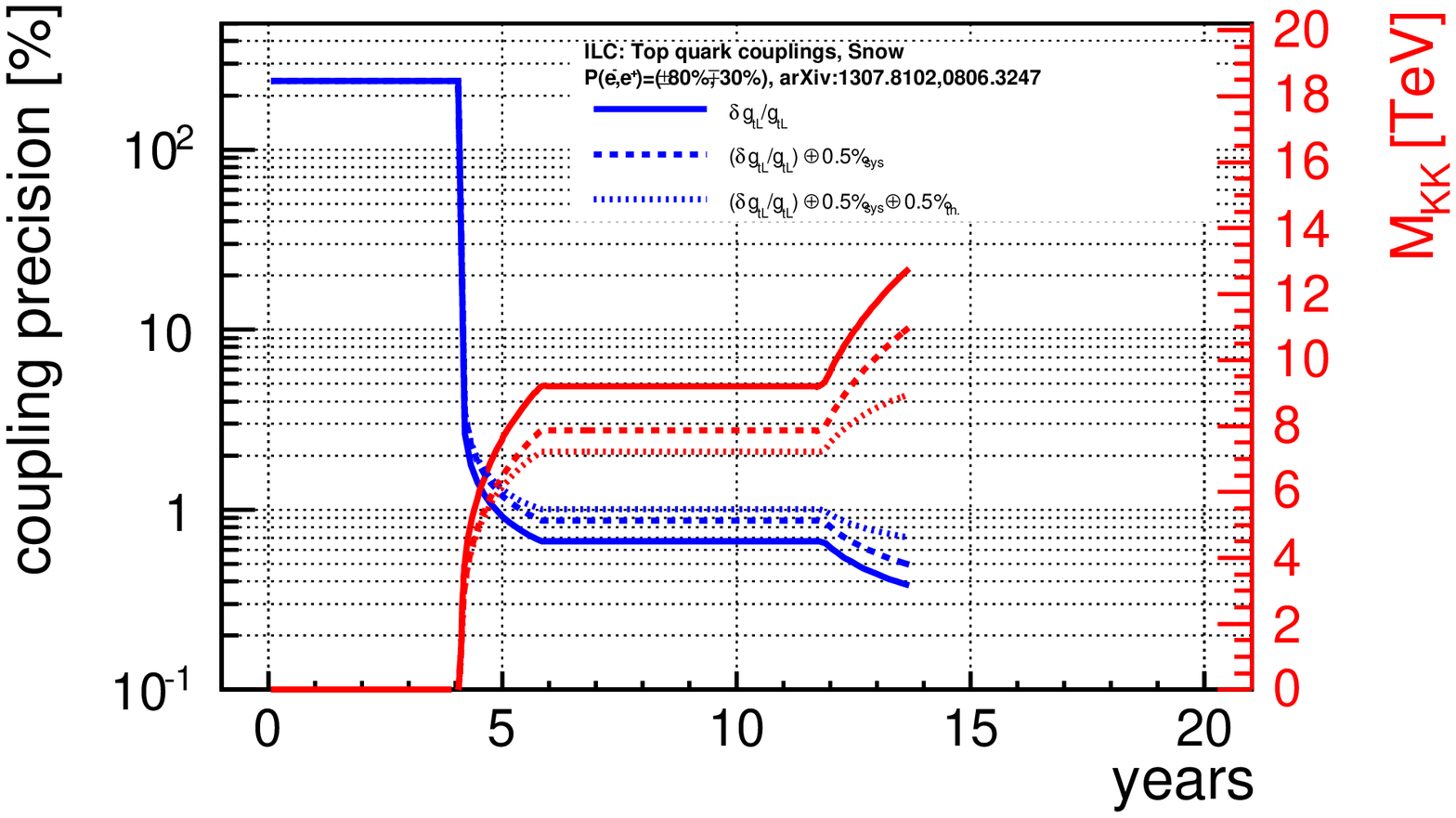}
  \caption{\color{myblue} Time-evolution of left-handed top coupling and derived from this the sensitivity to the mass scale of Kaluza-Klein excitations in an extra-dimension model~\cite{Richard:2014upa} for scenarios I-20 and Snow.\label{fig:topcoup_ISnow}}
\end{figure}

The precision measurement of the electroweak couplings of the top quark is a key item on the
ILC physics program. It requires beam polarisation in order to disentangle the couplings to the $Z$ boson and to the photon, which have different chiral properties. Besides being an important
test of the Standard Model in its own right, the top quark couplings are a prime indicator
for physics beyond the SM. Due to its uniquely large mass and thus its particularly strong coupling to the Higgs boson there is a strong motivation to expect that new phenomena would
become visible first in the top sector.

{\color{myblue}
Figures~\ref{fig:topcoup_GH} and~\ref{fig:topcoup_ISnow} show the time evolution expected for the left-handed top coupling based on~\cite{Amjad:2013tlv}. It also shows as an example the sensitivity to the mass scale of new physics in a Extra-Dimension model derived from excluding deviations of the left-handed top coupling from its Standard Model prediction~\cite{Richard:2014upa}. In this model, indirect sensitivity for new physics can extend easily into the $10-15$-TeV regime, which is beyond the reach of direct searches for resonances at the HL-LHC, which has been estimated to $5-6$\,TeV~\cite{Agashe:2013hma}. The measurement of the electroweak couplings of the top quark
requires at least $\sqrt{s}>450$\,GeV. 
}

\subsection{Higgs self-coupling}

As a final corner stone of the Higgs mechanism, a non-zero value of the Higgs self-coupling needs
to be established.  It has been claimed that a few per mille measurement of the $HZZ$ coupling can 
probe the Higgs self-coupling at the 30\% level through radiative corrections~\cite{McCullough:2013rea}.  However, an anomalous $HZZ$ coupling can arise from any number of sources, of which the Higgs self-coupling is but one.
An unambigous tree-level probe of the  Higgs self-coupling requires a measurement of the  double Higgs production cross section. At the High-Luminosity LHC, double Higgs production could eventually be observed
in the $\gamma\gamma b\bar{b}$ final state, however current fast simulation studies are rather pessimistic~\cite{ATLAS:Higgsself}. At the ILC, double Higgs production can be observed for $\sqrt{s}\ge 450$\,GeV,
but also here the measurement is challenging and requires a large amount of luminosity.

A detailed study based on full simulation of the ILD detector concept at $\sqrt{s}=500$\,GeV originally assuming $m_H=120$\,GeV~\cite{Tian2013:LCNote} has been updated recently~\cite{Duerig2014:privcom} to $m_H=125$\,GeV. Preliminary results have been obtained
for both unlike-sign helicity configurations of the beams, showing only a slight preference for
left-handed electrons and right-handed positrons. Combining the $HH \to b\bar{b}b\bar{b}$ and $HH \to b\bar{b}WW^*$ channels, a precision of $30\%$ has been demonstrated 
in full detector simulation assuming an integrated luminosity of $4$\,ab$^{-1}$, shared equally between $P(e^-e^+)=(\pm 80\%,\mp 30\%)$.
Recently, several opportunities to improve the sensitivity of the
analyses have been identified and are currently being implemented.
We include the estimated impact of these improvements in our plots.

\begin{figure}[htb]
\centering
\includegraphics[width=0.55\textwidth]{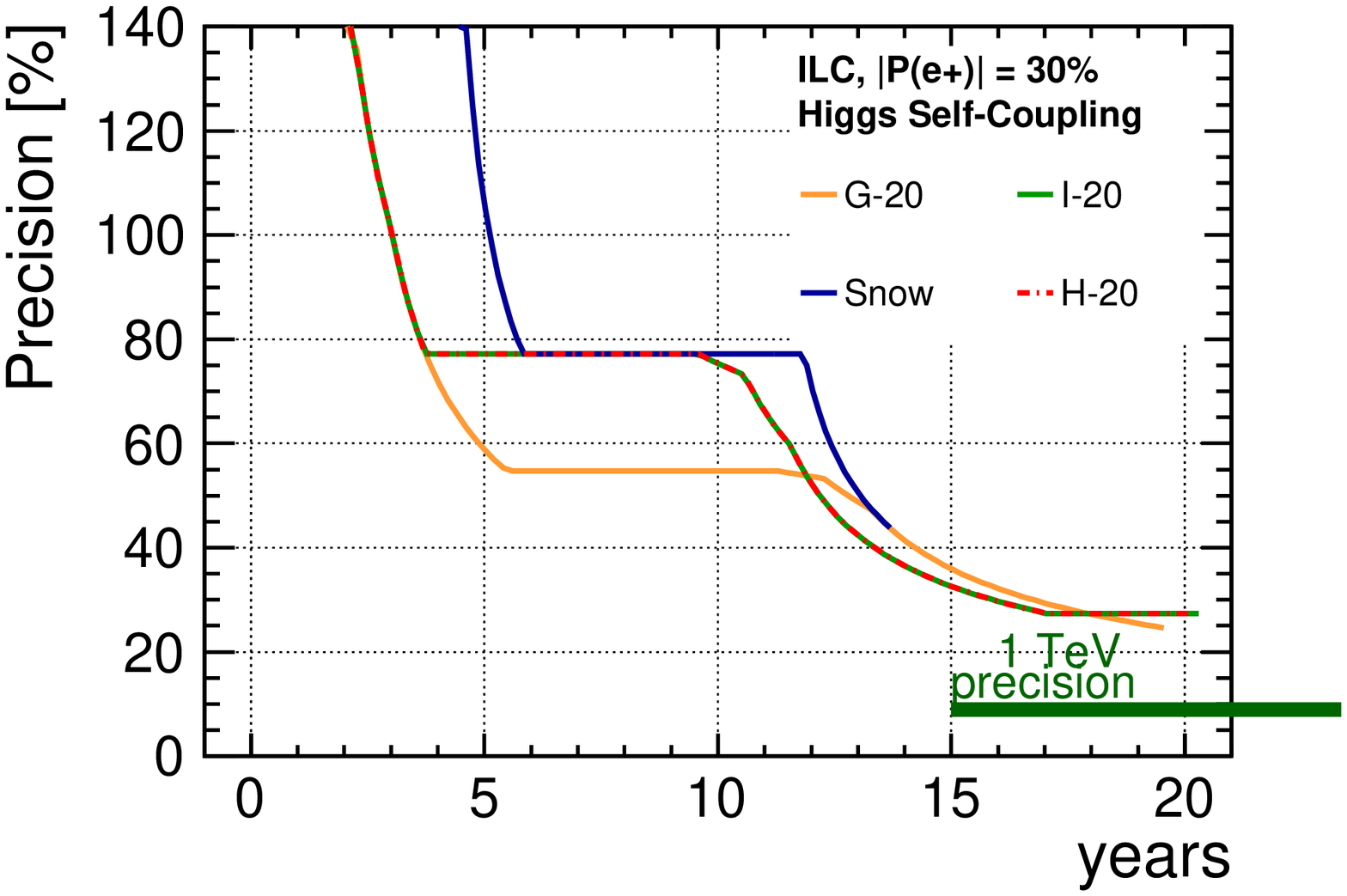}
\includegraphics[width=0.4\textwidth]{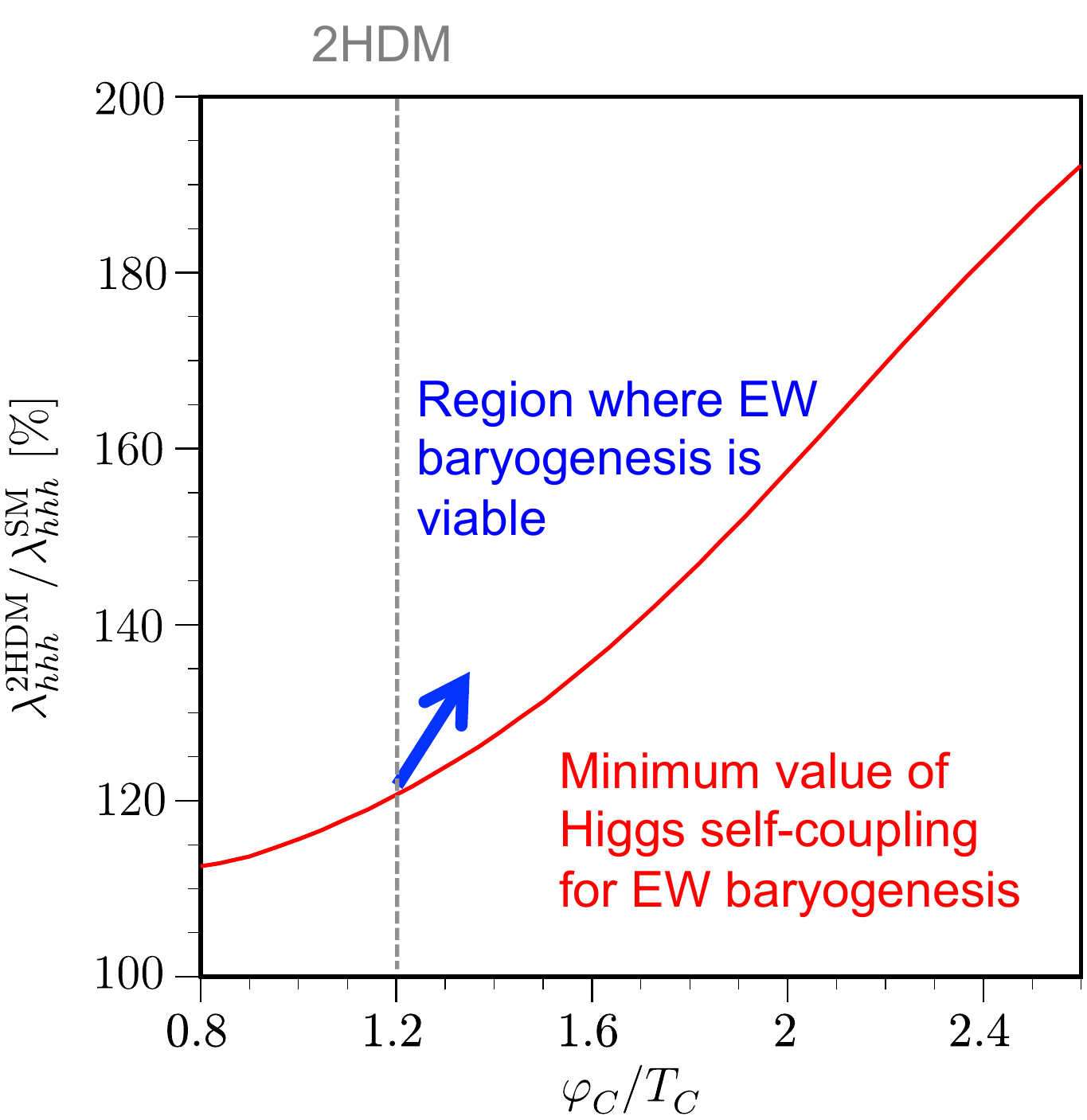}
  \caption{Left: Time evolution of the precision on the Higgs self-coupling for various running scenarios, based on the existing full simulation results exploiting the decay modes $H\to b\bar{b}$ and $H \to WW$, including an estimate of the effect of the ongoing analysis improvements
  (kinematic fitting, matrix element method, colour singlet jet clustering.) The green line shows for comparison the precision for the $1$\,TeV upgrade. Right: Example of allowed values of the Higgs self-coupling in Two-Higgs-Doublet-Models with electroweak baryogenesis normalised to the SM value~\cite{Kanemura:2012hr}. The minimal
  deviation is in the order of $20\%$, but the Higgs self-coupling could also be twice as high as in the SM. 
  \label{fig:ZHH}}
\end{figure}

The left-hand panel of figure~\ref{fig:ZHH} shows the time evolution of the precision on the
Higgs self-coupling for the scenarios G-20, H-20, I-20 and Snow. The helicities are chosen according to table~\ref{tab:pollumirel}. Before the luminosity upgrade, the precision is modest, but after the full H-20 program, a precision of $27\%$ can be reached~\cite{LCCPhysGroup:2015}. This would clearly demonstrate the existance of
the Higgs self-coupling.
The green line indicates for comparison the precision
that would be reached with the $1$\,TeV ILC upgrade
Here, a precision of $10\%$ or better can be achieved.

These numbers can be contrasted with expectations from various extensions of the Standard Model.
While only very small deviations are expected in the MSSM, deviations of 20\% or more can
be expected from models of electroweak baryogenesis~\cite{Kanemura:2012hr}, as shown in the right-hand panel of figure~\ref{fig:ZHH}.

{\color{myblue} In addition, it should be kept in mind that the double Higgs production mechanisms at the two center-of-mass energies are very different. In particular, the
sign of the interference term is different for double Higgsstrahlung and double Higgs production in $WW$-fusion. This means that a deviation of $\lambda$ from its Standard Model value will  lead to a larger
cross-section for one process and a smaller cross-section for the other. Thus both measurements are complementary in their sensitivity to new physics. }
 

\subsection{Natural supersymmetry: light higgsinos}
Among of the prime motivations to expect physics beyond the Standard Model is the so-called
hierarchy problem. It arises in the Standard Model since the mass of the Higgs boson as elementary scalar receives large corrections from quantum loops. Due to the large difference between the electroweak scale and the Planck scale, where gravity becomes strong, a fine-tuning of parameters to about $34$ digits would be required in order to keep the Higgs boson mass
close to the elctroweak scale, where it has been observed now by the LHC.

In Supersymmetry, the corresponding loop-diagrams with supersymmetric partner cancel the
corrections, up to a small rest depending on the mass differences between the SUSY particles
and their SM partners. If the masses of some SUSY particles become too large, again a certain
amount of fine-tuning creeps in, albeit at much smaller levels than in the pure SM. SUSY models
which try to minimize the amount of fine-tuning needed to stabilize the Higgs (and also the $Z$) boson mass have been titled ``Natural Supersymmetry''.

The most basic prediction of Natural SUSY models is that the lightest SUSY particles are
a triplett of Higgsinos, whose mass is given by the non-SUSY-breaking parameter $\mu$.
The mass splitting within the triplet is inversely proportional to the SUSY-breaking gaugino mass parameter $m_{1/2}$. Coloured SUSY particles can be much heavier, with the $\tilde{t}_1$
mass between $1$ and $2$\,TeV and the gluino mass in the $1.5-5$\,TeV range~\cite{Mustafayev:2014lqa}.

\begin{figure}[htb]
\centering
\includegraphics[width=0.395\textwidth]{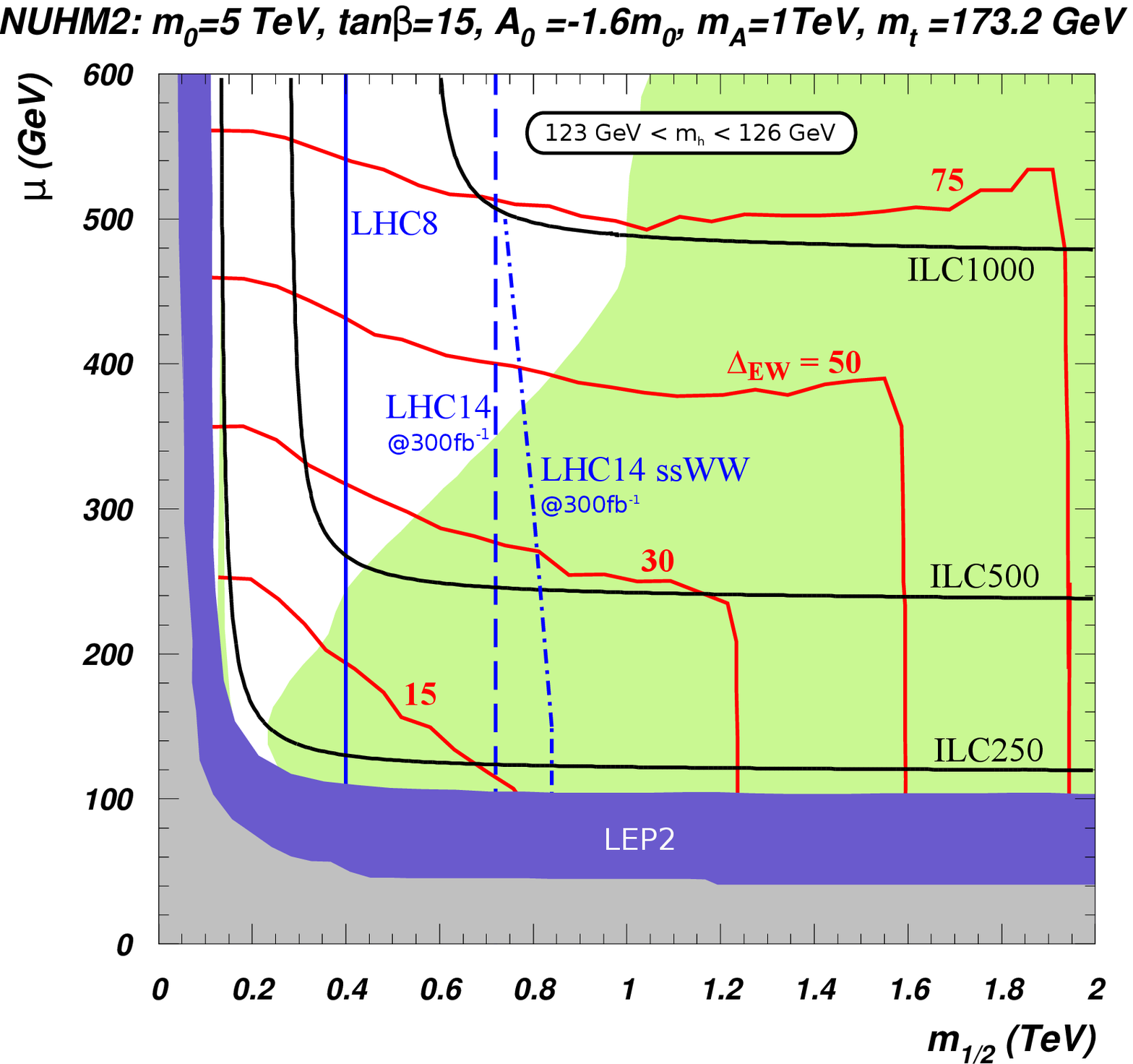}
\hspace{0.1cm}
\includegraphics[width=0.55\textwidth]{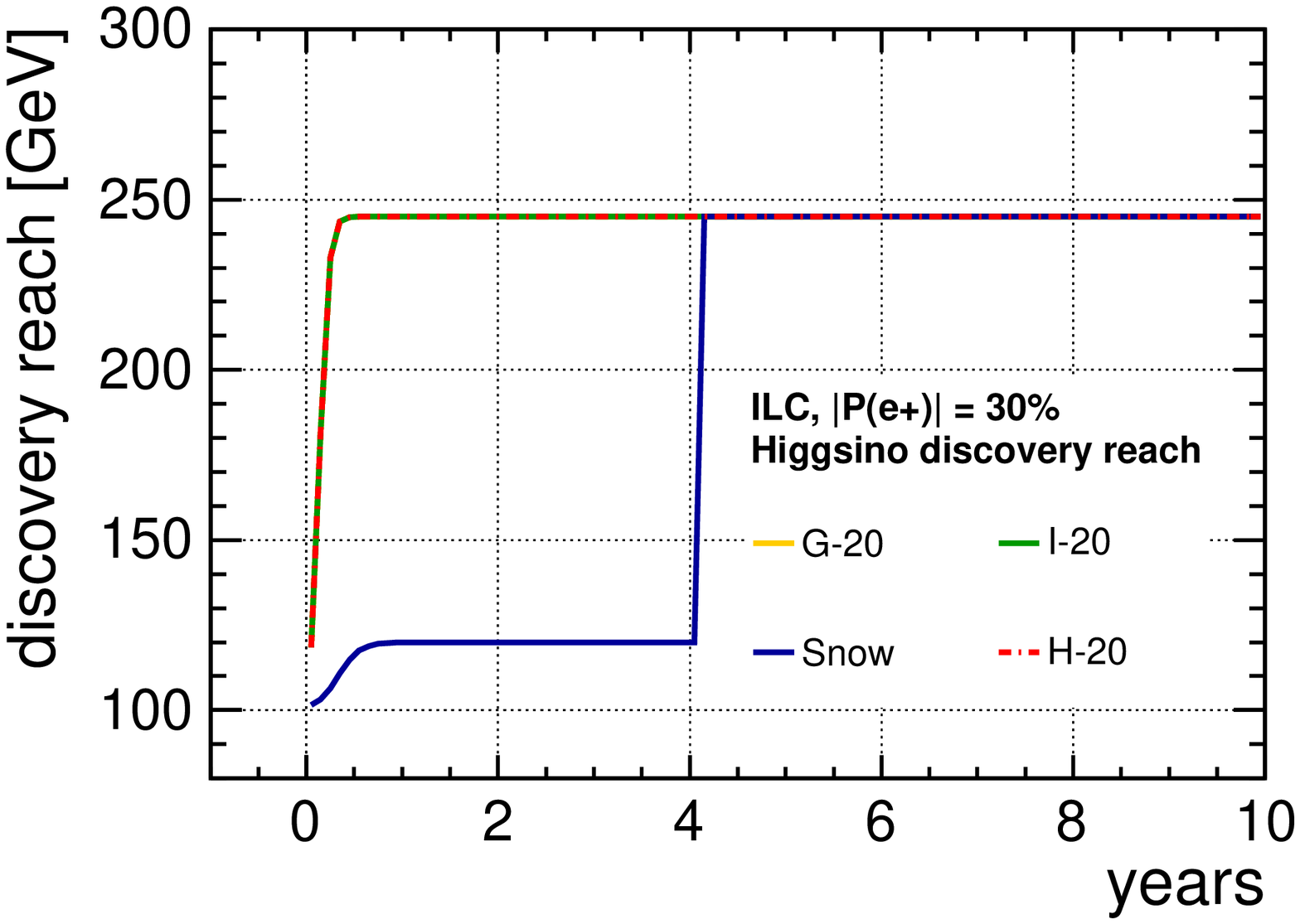}
  \caption{Left: Discovery reach of LHC and ILC in the $\mu$ vs $m_{1/2}$ plane. The ILC covers
   substantial parameter space with low fine-tuning inaccessible to the LHC. Further description see text. From~\cite{Baer:2013xua}.
  Right: Time evolution of Higgsino discovery reach in different ILC running scenarios.
  \label{fig:higgsinos}}
\end{figure}

It has been shown that these light Higgsinos can be observed at the ILC, including precise 
measurements of their properties~\cite{Berggren:2013vfa}. At the LHC with $\sqrt{s}=14$\,TeV, they can be easily observed in cascade decays if the gluino is light enough, i.e. $m_{1/2} \lesssim 0.7$\,TeV assuming gaugino mass unification, c.f. the left part of figure~\ref{fig:higgsinos}. This can be extended somewhat by same-sign dilepton searches to about
$0.9$\,TeV.
In constrast, the range of the ILC is nearly independent of $m_{1/2}$, actually in the benchmark points studied in~\cite{Berggren:2013vfa} $m_{1/2}$ was approximately $5$\,TeV.
The green shaded area indicates the region of parameter space in which the Dark Matter relic density is not larger than the observed value, while the read lines indicate the degree of fine-tuning (the lower $\Delta_{EW}$ the less fine-tuning). With $\sqrt{s}=500$\,GeV, the ILC and LHC14 together
cover nearly all the region with a fine-tuning less than $30$, while ILC with$\sqrt{s}=1$\,TeV
probes the region up to $\Delta_{EW}=75$.

At the ILC, Higgsinos would be discovered very quickly once they are kinematically accessible.
This is illustrated in the right part, which shows the discovery reach as a function of time
for our running scenarios. With the
integrated luminosities of the full ILC programm, the Higgsinos masses and cross-sections
could then be measured to the level of $1\%$ or better, which enables to prove that the discovered particles are indeed Higgsinos, to determine the parameters of the underlying model
and e.g. to constrain the gaugino mass parameters even if they are in the multi-TeV regime. We point out that polarised beams
are essential for this enterprise.  

Obviously, reaching the highest possible center-of-mass energy as early as possible is 
desirable in terms of discovery potential of the machine, not only in case of Higgsinos,
but also in view of other electroweak states with small mass differences which could
espace detection at the LHC.

\subsection{WIMP Dark Matter}
\label{subsec:physWIMP}

\begin{figure}[htb]
\centering
\includegraphics[width=0.45\textwidth]{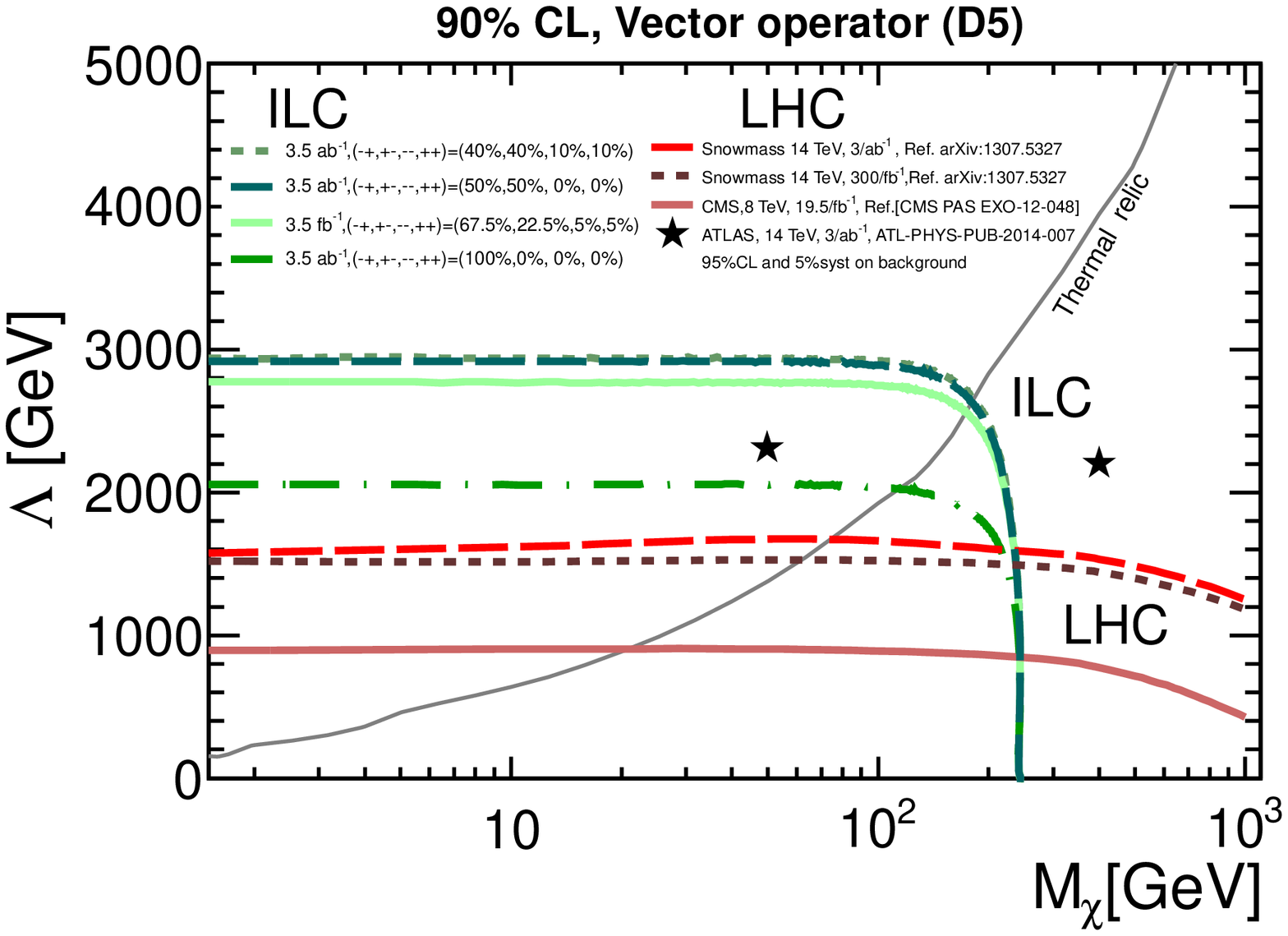}
\hspace{0.1cm}
\includegraphics[width=0.475\textwidth]{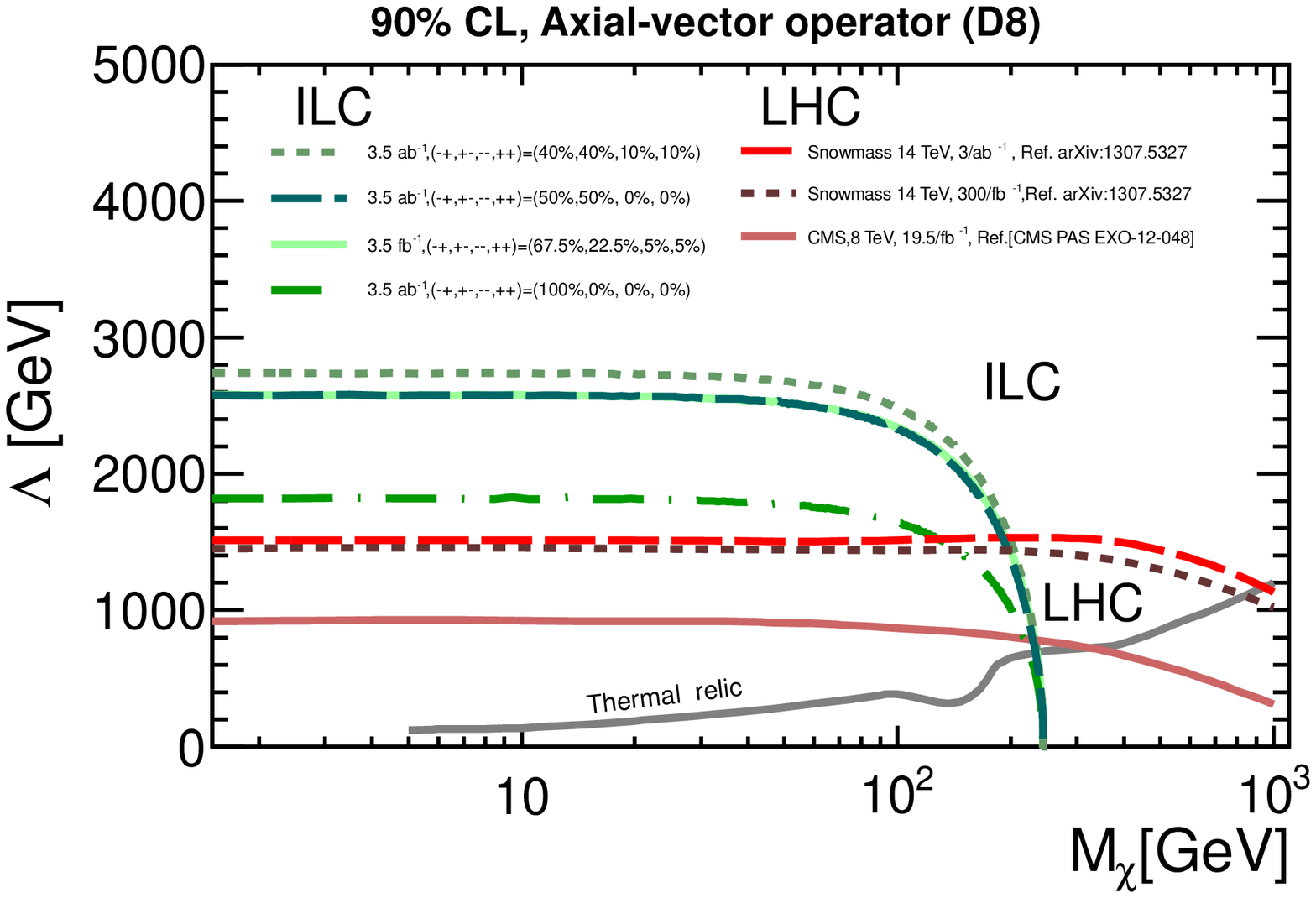}
  \caption{$90\%$ CL reach for WIMP dark matter at LHC and ILC in the plane of effective
  operator scale vs WIMP mass.
  Left: For a vector operator mediating the WIMP interaction.
  Right: For an axial-vector operator mediating the WIMP interaction.
  \label{fig:WIMP_LHC_ILC}}
\end{figure}

One of the prime tasks of current and future colliders is to identify the nature of Dark Matter. WIMP Dark Matter can be searched for at colliders in a rather model-independent
manner by looking for mono-jet or mono-photon events. Figure~\ref{fig:WIMP_LHC_ILC} 
compares actual LHC results, LHC projections and ILC projections. For the ILC, an
integrated luminosity of $3.5$\,ab$^{-1}$ is assumed at $\sqrt{s}=500$\,GeV, and the
result is shown for different polarisation sharings. Note that data-taking with like-sign
polarisation configurations is important in particular for the case of an axial-vector-type
of interaction between the WIMPs and SM particles. For WIMP masses well below the kinematic
limit, the reach in the effective operator scale (thus the scale of new physics) is basically
independent of the WIMP mass.

\begin{figure}[htb]
\centering
\includegraphics[width=0.45\textwidth]{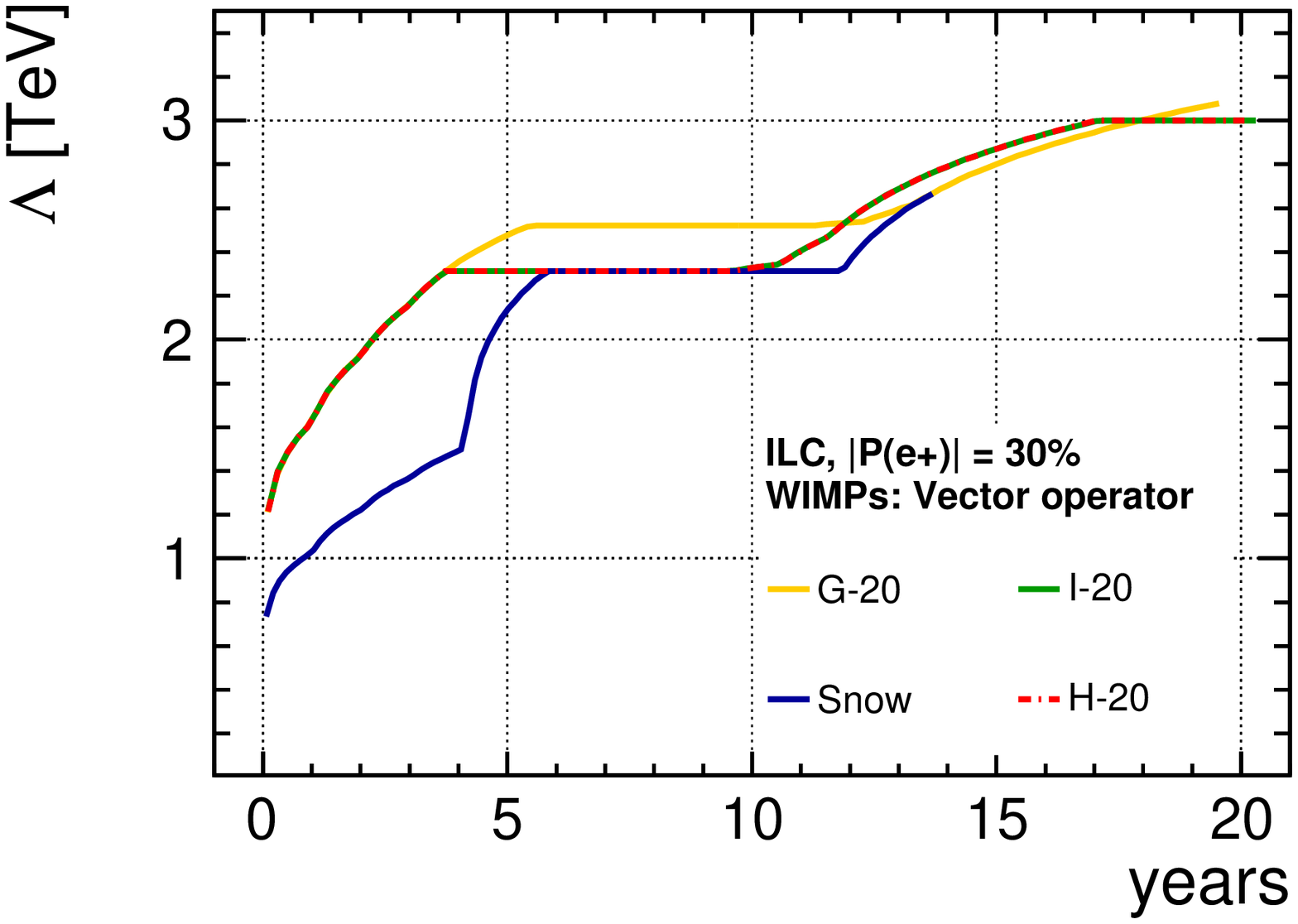}
\hspace{0.1cm}
\includegraphics[width=0.45\textwidth]{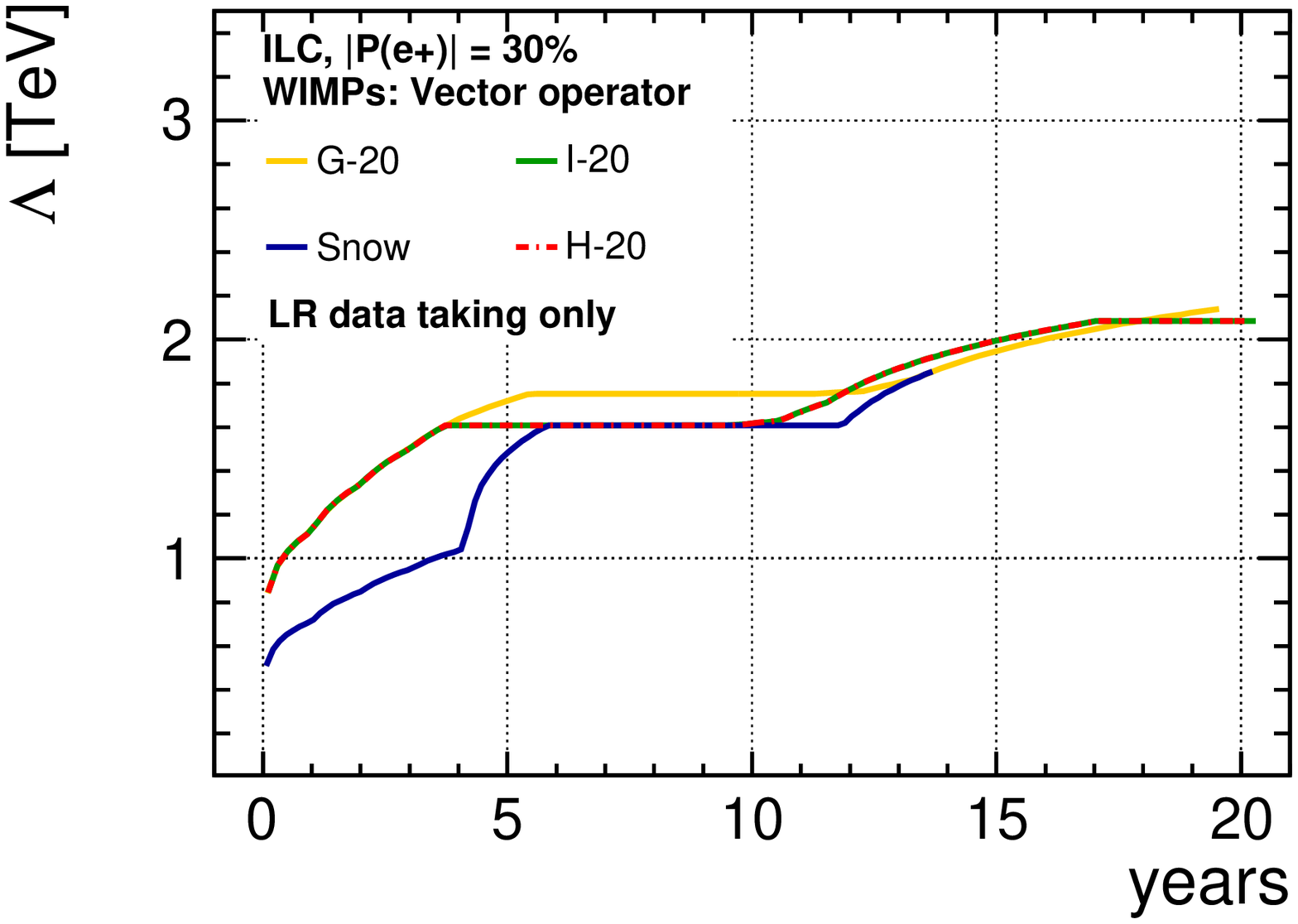}
  \caption{$90\%$ CL reach for WIMP dark matter at ILC for a WIMP mass of $10$\,GeV
  in the vector operator case in our running scenarios.
  Left: Polarisation sharing as proposed in table~\ref{tab:pollumirel}
  Right: All data taken with $P(e^-,e^+) = (-80\%,+30\%)$.}
  \label{fig:WIMP_running}
\end{figure}

Figure~\ref{fig:WIMP_running} shows the time evolution of the reach in new physics scale
$\Lambda$ in our running scenarios, taking as example the vector operator case and a WIMP
mass of $10$\,GeV. Again it can be seen easily that early operation at high center-of-mass 
energies is preferred. {\color{myblue} The right panel of figure~\ref{fig:WIMP_running} illustrates
the loss in sensitivity if all data were collected exclusively with $P(e^-,e^+) = (-80\%,+30\%)$.}

%% file: tth.tex
\section{Maximum Centre-of-Mass Energy Reach of $\sim$500 GeV ILC and the Top Yukawa Coupling}
\label{sec:tth}

The top Yukawa coupling is measured at the ILC
from the process $e^+e^- \rightarrow t \overline{t} h$, which
opens kinematically at around $\sqrt{s}=475$\,GeV.
Full detector simulation studies showed that at $\sqrt{s}=500$\,GeV,
the top Yukawa coupling can be determined with a precision of $9.9\%$ based
on an integrated luminosity of $1$\,ab$^{-1}$ with $P(e^-,e^+)=(-80\%,+30\%)$~\cite{Yonamine:2011jg}.
In terms of the running scenarios proposed in this document, this translates
into final precisions between $5\%$ and $7\%$, c.f.~Figures~\ref{fig:HiggsCouplingsHbbHtt} and~\ref{fig:HiggsCouplingsHbbHttlin}.

\begin{figure}[htb]
\centering
\includegraphics[width=0.8\textwidth]{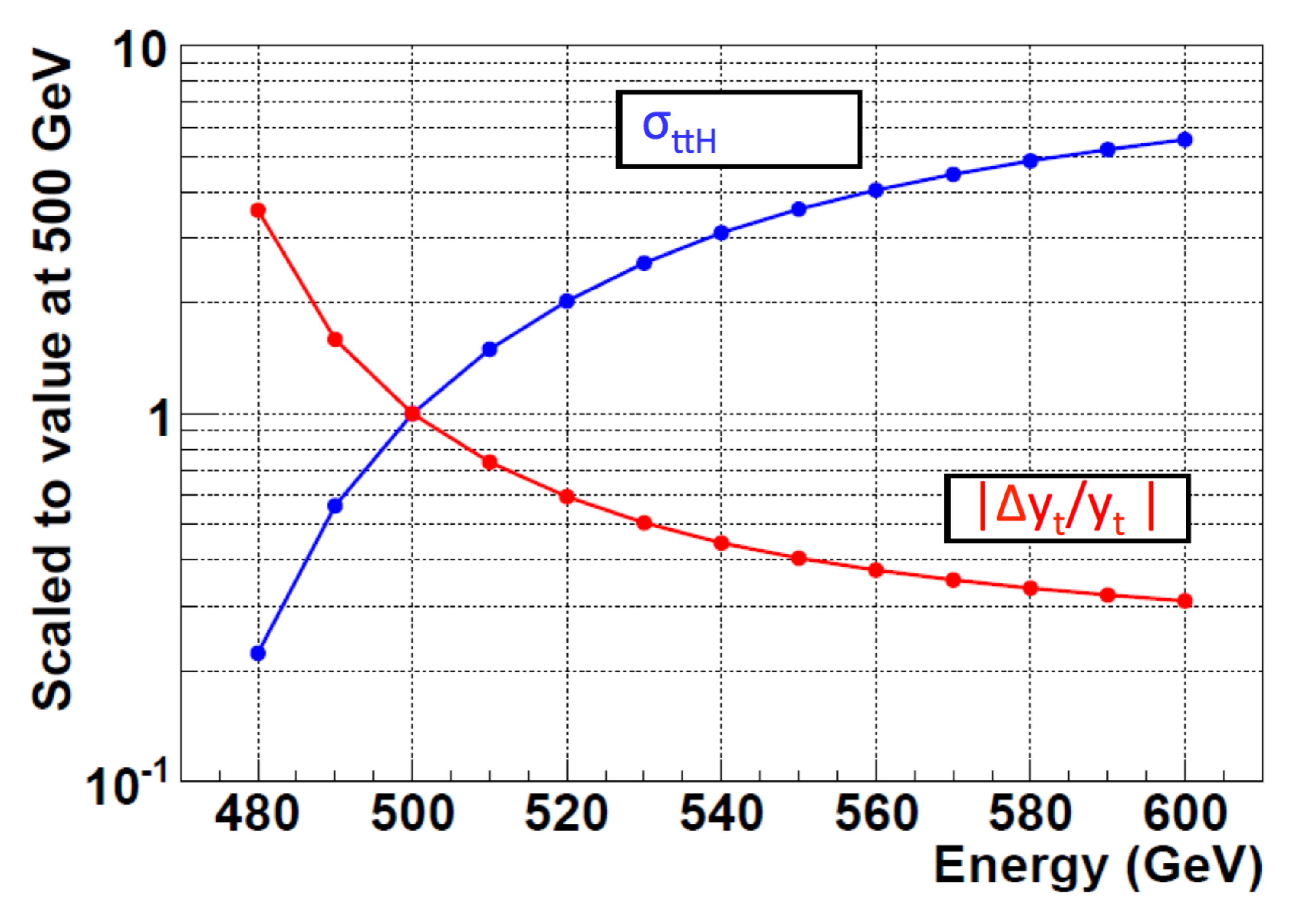}
  \caption{Relative cross section and top Yukawa coupling precision versus centre-of-mass energy,
  extrapolated based on scaling of signal and main background cross-sections. 
\label{fig:tth_480}}
\end{figure}

Figure \ref{fig:tth_480} shows the relative cross section for $t\bar{t}h$ production as
a function of $\sqrt{s}$, which shows that it is still steeply rising at $\sqrt{s}=500$\,GeV,
increasing nearly four-fold by $\sqrt{s}=550$\,GeV. Since at the same time the main
backgrounds, e.g. from non-resonant $tbW$ and $t\bar{t}b\bar{b}$ production, decrease, the 
precision on the top Yukawa coupling improves by better than a factor of two w.r.t. 
$\sqrt{s}=500$\,GeV for the same integrated luminosity. This significant improvement in 
the important top Yukawa coupling
parameter motivates serious consideration of extending the upper
center-of-mass reach of the nominally $500$\,GeV ILC to about $550$\,GeV.

On the other hand it should be noted that for $\sqrt{s}<500$\,GeV the
cross-section drops quickly. For $\sqrt{s}=485$\,GeV, a reduction of $3\%$ in $\sqrt{s}$, 
the uncertainty of the top Yukawa would be twice as large as at $\sqrt{s}=500$\,GeV. 
Thus reaching at least $\sqrt{s}=500$\,GeV is essential to be
able to perform a meaningful measurement of the top Yukawa coupling.

%% file: additional.tex
%
%
%
\section{Modifications of Running Scenarios in Case of New Physics}
\label{sec:modNP}
The above running scenarios have been derived based on the particles we know today.
However there are many good reasons to expect discoveries of new phenomena at the
LHC or the ILC itself. Obviously, such a discovery would lead to modifications of
the proposed running scenarios. Since the possibilities are manifold, we outline here
the basic techniques which exploit the tunability of centre-of-mass energy and beam
helicities at the ILC in order to characterize new particles. 
In practice, the inital run at $\sqrt{s}=500$\,GeV would serve as a scouting run. 
Careful analysis of these data will then give some first information, which would guide
the optimisation of the running program for the following years.

\subsubsection*{Threshold Scans} 
As in the case of the top quark or the $W$ boson, threshold scans are also important tools
for a precise determination of the masses of new particles. This has been studied in the
literature mainly for the example of supersymmetry, in particular for sleptons, but also
for charginos and neutralinos. In the case of e.g.\ a smuon, a threshold scan with a moderate
integrated luminosity of $90$\,fb$^{-1}$ yields a mass measurement of the smuon to better than $0.2$\%, while $500$\,fb$^{-1}$ with $P(e^-,e^+)=(+80\%,-30\%)$ at $\sqrt{s}=500$\,GeV
yields an uncertainty which is about a factor of $2.5$ larger~\cite{Berggren:smuons}.

In our scenario H-20, for instance, the mass of the right-handed smuon would be known to about $800$\,MeV from the $200$\,fb$^{-1}$ taken with this polarisation during the first $4$ years of 
ILC operation. This would set the energy points to scan at a later time in order
to improve the precision by about a factor of $4$ within a few months of running. It should
be noted that depending on the exact masses, the data from such a scan can also be interesting
for various other physics. Similar improvements are expected for scans of neutralino and
chargino thresholds. The importance of the gain in precision for the determination of the
parameters of the underlying model has recently be highlighted e.g.\ in~\cite{aoife}.

Even more fundamentally, the shape of the threshold allows an unambiguos determination of the
spin of the produced particles~\cite{Choi:2001ww, Choi:2002mc}. This is an essential ingredient
to demonstrating - or falsifying -  the supersymmetric nature of the new particle! 

\subsubsection*{Operation with Different Beam Helicities} 
Once a new particle has been discovered, it is not sufficient to determine its mass. 
Also its couplings, and especially their chiral properties need to be investigated. The
key to this enterprise are the polarised beams of the ILC: whether it is supersymmetry or
some other kind of new physics, the measurement of all four polarised cross sections
will give unique insights to the nature of the new physics.

An example is the case of generic WIMP dark matter from section~\ref{subsec:physWIMP}, which could be observed and characterized
in a mono-photon type of analysis via the energy spectrum of the associated initial state radiation, and which has been studied e.g.\ in~\cite{Bartels:2012ex, Choi:2015zka}. In such a case,
it could become important to take more data in the like-sign helicity combinations in order
to obtain the full picture of the chiral couplings for a model-independent analysis.

But also in supersymmetry there are interesting cases where like-sign helicity data will give
decisive clues: For instance the mixed production of left- and right-handed selectrons in
$e^+_L e^-_L \to \tilde{e}^+_R \tilde{e}^-_L$ and $e^+_R e^-_R \to \tilde{e}^+_L \tilde{e}^-_R$
proceeds exclusively via $t$-channel neutralino exchange. Since the higgsino component has
an extremely small contribution due to the tiny Yukawa coupling of the electron, the measurements of these cross sections will give important insights into the mixing nature of the neutralinos~\cite{MoortgatPick:2005cw}. This in turn is a key ingredient on predicting e.g.\ the dark matter relic density in order to identify the LSP as the dark matter particle.

\subsubsection*{Operation at Specific Centre-of-Mass Energies}
In many cases, $500$\,GeV will not be the optimal center-of-mass energy to scrutinize
a specific process. For instance in all cases which rely on the recoil of an invisible (dark matter) or nearly invisible system against initial state radiation, data taking near the
threshold leads to an improved recoil mass resolution and to a larger dependency of the
cross section on model parameters. This is in particular true for the case of light, near
degenerate higgsinos~\cite{Berggren:2013vfa}, as they are predicted in natural supersymmetry.

Another example occurs in case of several new particles with similar final state products or
complicated decay chains. In such cases, when new physics becomes a background to itself, the ability to tune the centre-of-mass energy so that only a part of the spectrum 
is kinematically accessible can be the key to precision measurements. An example is the
cross section measurement of $e^+e^- \to \tilde{\tau}_1 \tilde{\tau}_2$, which gives a handle
on the mixing in the $\tilde{\tau}$ sector: if $e^+e^- \to \tilde{\tau}_2 \tilde{\tau}_2$
is kinematically allowed, the mixed process is nearly impossible to fish out of the pair production
background, obstructing this important measurement~\cite{Bechtle:2009em}. 

As opposed to threshold scans, these close-to-threshold operation points are not required to
be exactly at the threshold. Thus, suitable energies can be identified in view of the whole
physics program, so that also these energies are useful for many different measurements.
In the higgsino case~\cite{Berggren:2013vfa}, for instance, the close-to-threshold energy
has been set to $350$\,GeV. While H-20 forsees only very little luminosity at $350$\,GeV, sufficient for the top threshold scan, more data at this energy can also be used efficiently for a large
variety of Higgs measurements. Thus, new physics will not neccessarily require additional single-purpose runs, but will primarily lead to a reoptimisation of the whole running program,
shifting the choices suggested here from today's perspective.

\section{Other Centre-of-Mass Energies}
\label{sec:other}

%
%
\subsubsection*{$\sqrt{s}=1$\,TeV}
\label{sec:1TeV}
As already pointed out above in case of the Higgs self-coupling, an extension
of the ILC to $\sqrt{s}=1$\,TeV offers significant improvements for many Standard
Model precision measurements including the Higgs boson properties. In addition it
immensely enlarges the reach of the ILC for direct production of new particles, 
in particular Dark Matter candidates. Therefore, the compatibility of the 
ILC baseline design with a later upgrade to $1$\,TeV should be preserved.

\subsubsection*{WW-threshold, Z-pole for physics, Z-pole for calibration}
\label{sec:lowenergy}
Beyond the physics goals in the 250-500 GeV energy range that have primarily been
the focus of the discussion in the previous sections, lower energy operation of the ILC must be considered for a number of important purposes, also discussed in~\cite{Baak:2013fwa}:
\begin{enumerate}
\item Precision measurements of the W boson mass at the W pair threshold
near $\sqrt{s}$ = 161 GeV could reach the few MeV regime. This measurement is qualitatively 
unique since it is subject to orthogonal experimental systematic uncertainties compared
to kinematic reconstruction of the $W$ mass in the continuum, and since it has a direct
theoretical interpretation.
\item Ultra-precise measurement of the left-right asymmetry ($A_{LR}$)
in the production of $Z$ bosons at $\sqrt{s} = M_Z$ based on $10^9$ $Z$ decays
(so-called Giga-$Z$)
offers an important update of the SLD measurements of $A_{LR}$ and
its tension with the LEP forward-backward asymmetry measurements.
\item Operation of the ILC at $\sqrt{s} = M_Z$ also offers a valuable 
calibration tool even at lower luminosities than those required for the Giga-$Z$
measurement. However SiD and ILD should quantify their needs more precisely.
\end{enumerate}
We note that optimal luminosity perfomance
at $\sqrt{s}<200$\,GeV will require a machine upgrade from the TDR design.

\subsubsection*{Target Integrated Luminosities and Polarisation Splitting}

Table~\ref{tab:lumiabstot1TeV} shows proposed total target integrated luminosities for further
 options of the ILC program, 
 in particular 
 precision measurements at the $Z$
 pole and a $W$ pair production threshold scan for a precision measurement of the $W$ mass. 
 The luminosities given here serve as guideline for future physics studies of these options. 
 Preliminary estimates indicate that a threshold scan with an integrated luminosity of 
 $500$\,fb$^{-1}$ would offer the potential to determine $m_W$ to the level of $1-2$\,MeV.
 $100$\,fb$^{-1}$ at the $Z$ pole corresponds to about $3\times10^9$ hadronic $Z$ events,
 thus corresponds to the full GigaZ programme. We stress however that also a ten times or even a hundred
 times smaller number of polarised $Z$ pole events would advance significantly our understanding of $A_{LR}$ 
 compared to the SLD measurement. Detailed requirements on the beam parameters for both centre-of-mass
energies should be worked out in the future.

\begin{table}[h]
\centering
  \renewcommand{\arraystretch}{1.10}
\begin{tabularx}{\textwidth}{*{4}{>{\centering\arraybackslash}X}} 
\hline
$\sqrt{s}$  &    
90\,GeV & 160\,GeV  \\
\hline
 $\int{\mathcal{L} dt}$ [fb$^{-1}$]           & 
 100       &  500  \\
\hline
\end{tabularx}
\caption{Proposed total target integrated luminosities for other $\sqrt{s}$. }
\label{tab:lumiabstot1TeV} 
\end{table}

Tables~\ref{tab:pollumirel1TeV} suggests a luminosity
sharing between the beam helicity configurations for these energies, while  and~\ref{tab:pollumiabs1TeV} gives the corresponding absolute integrated luminosities
per helicity configuration.

\begin{table}[h]
\centering
  \renewcommand{\arraystretch}{1.10}
\begin{tabularx}{\textwidth}{*{5}{>{\centering\arraybackslash}X}} 
\hline
        & \multicolumn{4}{c}{fraction with $\operatorname{sgn}(P(e^-),P(e^+))= $ } \\
           & (-,+) & (+,-) & (-,-) & (+,+) \\
\hline
$\sqrt{s}$ & [\%]  &  [\%] & [\%]  & [\%]  \\ 
\hline
1\,TeV     &  40  &  40   &  10   &  10   \\
90\,GeV    &  40  &  40   &  10   &  10   \\
160\,GeV   & 67.5 &  22.5 &  5    &   5   \\
\hline
\end{tabularx}
\caption{Relative sharing between beam helicity configurations proposed for low energy and $1$\,TeV running.}
\label{tab:pollumirel1TeV} 
\end{table}

\begin{table}[h]
\centering
  \renewcommand{\arraystretch}{1.10}
\begin{tabularx}{\textwidth}{*{5}{>{\centering\arraybackslash}X}}    
\hline
        &  \multicolumn{4}{c}{integrated luminosity with $\operatorname{sgn}(P(e^-),P(e^+))= $ } \\
           & (-,+)       & (+,-)       & (-,-)       &  (+,+)     \\
\hline
$\sqrt{s}$ & [fb$^{-1}$] & [fb$^{-1}$] &  [fb$^{-1}$] & [fb$^{-1}$] \\ 
\hline
1\,TeV      &  3200   	 & 3200        &  800	      &   800  \\
90\,GeV     &    40   	 &   40        &   10	      &    10  \\
160\,GeV    &   340   	 &  110        &   25	      &    25  \\
\hline
\end{tabularx}
\caption{Integrated luminosities per beam helicity configuration resulting from the fractions in table~\ref{tab:pollumirel1TeV}.}
\label{tab:pollumiabs1TeV} 
\end{table}

The exact priority of the low energy runs will largely depend on the future results
of the LHC and the first round of ILC operation. The longer a direct discovery of
new physics evades experimental proof, the more relevant ultra-precise measurements
of the most fundamental parameters of the Standard Model will become.

%% file: conclusions.tex
\section{Conclusions}
\label{sec:concl}

This report summarizes studies of possible operating scenarios for the
500 GeV ILC, the collider describing in the ILC TDR.
The preferred scenario is H-20.
After starting operation at the full
centre-of-mass energy of $500$\,GeV, 
running is planned at $250$ and $350$\,GeV before
the collider luminosity is upgraded for intense
running at $500$\,GeV and at $250$\,GeV.
This scenario (H-20) optimizes the  possibility of discoveries of new physics while 
making the earliest measurements of the important Higgs properties. It includes
a sizable amount of data taken at $\sqrt{s}=250$\,GeV, since based on current knowledge 
this is the only proven way to guarantee a fully model-independent precision determination
of the Higgs mass and its coupling to the $Z$ boson.

We note the physics impact of the ILC is significantly improved if the
maximum energy of the $\sim 500$\,GeV ILC is stretched to $\sim 550$\,GeV where the
top Yukawa precision is more than a factor of two times better than at $500$\,GeV.

This report emphasizes the physics that we are absolutely certain will be done with the ILC and the
   operational accelerator plans for achieving the best outcomes for that physics.
   This physics includes precision measurements of the Higgs boson, the top quark, and possibly measurements
   of the W and Z gauge bosons.  While this certain program provides a compelling and impactful scientific outcome,
   discoveries by the LHC or the early running of the ILC could expand the scientific impact
   of the ILC even further.  There are existing scientific motivations to anticipate such possibilities.
   Such discoveries could alter the run plan from that described by H-20,
    as operations at our near the threshold of a pair-produced new particle, for example,
   would be added, a capability that is one of the particular operational strengths of the ILC.

%% file: higgsMeasurements.tex
\appendix
\section{Input Precisions to the Higgs Coupling Fit}
\label{sec:higgsmeas}

The accuracies for the  cross section and $\sigma\cdot BR$ measurements 
used  in the Higgs coupling fits for this paper are summarized in Table~\ref{tab:stimesbrbase}
for the reference luminosities and energies used in the actual simulation studies. Many of the results were obtained from the Snowmass ILC Higgs White Paper~\cite{Asner:2013psa}. Since Snowmass 2013 Higgstrahlung cross section measurement using hadronic Z decays have been developed for  $\sqrt{s}=250$\,GeV~\cite{Tomita2015:hadronrecoil}
and $\sqrt{s}=350$\,GeV~\cite{Thomson2014:hadronrecoil}, and many studies discussed in the Snowmass ILC Higgs White Paper have 
been reanalyzed for $\sqrt{s}=350$\,GeV and/or a Higgs mas of $125$\,GeV. All analyses have been performed assuming a beam polarisation of $P(e^-,e^+)=(-80\%,+30\%)$. A snapshot of all updates at the time of writing of this report can be found in~\cite{Tian2015:higgssummary}. Results for the opposite combination, $P(e^-,e^+)=(+80\%,-30\%)$
have been obtained by scaling factors derived from polarisation dependence of the signal and main background processes. Typically, they are $1$ for $ZH$ production, while for $WW$ fusion signals
a factor $4$ is  assumed. The limits on invisible Higgs have been analysed for both
helicities, and $P(e^-,e^+)=(+80\%,-30\%)$ has been found to yield more stringent limits, 
corresponding to a scaling factor of $0.7$. For the simulation study of $t\bar{t}h$ production, $P(e^-,e^+)=(+80\%,-30\%)$
has been found to perform worse than the reference polarisation $P(e^-,e^+)=(-80\%,+30\%)$ by a factor of $1.36$.
Within the global fit, all measurements are scaled to the desired integrated luminosities and
polarisation sharings before extracting the coupling precisions for a given combination of datasets.


\begin{table}[htb]
 \begin{center}
 \begin{tabular}{|l|r|r|r|r|r|r|r|}
   \hline
  $\int{\mathcal{L}}dt$ at $\sqrt{s}$ & \multicolumn{2}{c|}{250\,fb$^{-1}$ at 250\,GeV} 
                                         & \multicolumn{2}{c|}{330\,fb$^{-1}$ at 350\,GeV }
                                         & \multicolumn{3}{c|}{500\,fb$^{-1}$ at 500\,GeV} \\
   \hline
  $P(e^-,e^+)$ & \multicolumn{7}{c|}{(-80\%,+30\%)}  \\

  \hline
production& $Zh$     & $\nu\bar{\nu}h$ & $Zh$     & $\nu\bar{\nu}h$        & $Zh$     & $\nu\bar{\nu}h$ & $t\bar{t}h$ \\
   \hline\hline
  $\Delta \sigma / \sigma$ & \cite{recoil250}~~2.0\% & - & \cite{recoil350, Thomson2014:hadronrecoil}~~1.6\% & - & 3.0 & - & - \\ \hline
   BR(invis.)~\cite{Ishikawa2014:LCWS}  & $<0.9\%$ & - & $<1.2\%$ & - & $<2.4\%$ & - & - \\ \hline \hline
   decay & \multicolumn{7}{c|}{$\Delta (\sigma \cdot BR) / (\sigma \cdot BR)$}  \\
  \hline
   $h \to b\bar{b}$              & 1.2\%    & 10.5\%          & 1.3\%    & 1.3\%                  & 1.8\%    & 0.7\%           & 28\%        \\
   $h \to c\bar{c}$              & 8.3\%    & -               & 9.9\%    & 13\%                   & 13\%     & 6.2\%           & -           \\
   $h \to gg$                    & 7.0\%    & -               & 7.3\%    & 8.6\%                  & 11\%     & 4.1\%           & -           \\
   $h \to WW^*$                  & 6.4\%    & -               & 6.8\%    & 5.0\%                  & 9.2\%    & 2.4\%           & -           \\
   $h \to \tau^+\tau^-$ &\cite{Kawada2015:ALWC}~~3.2\% & - &\cite{Kawada2015:extra}~~3.5\% & 19\% & 5.4\%    & 9.0\%           & -           \\
   $h \to ZZ^*$                  & 19\%     & -               & 22\%     & 17\%                   & 25\%     & 8.2\%           & -           \\
   $h \to \gamma\gamma$          & 34\%     & -               & 34\%     &\cite{Calancha2013:LCWS}~~39\% & 34\% &\cite{Calancha2013:LCWS}~~19\% & - \\
   $h \to \mu^+\mu^-$~\cite{Hmumu}            & 72\%     & -               & 76\%     & 140\%                  & 88\%     & 72\%            & -           \\
   \hline
  \end{tabular}
  \caption{Expected accuracies for cross section and cross section times branching ratio
measurements for the $125\,$GeV Higgs boson as provided as input to the coupling fit. All values obtained 
from full detector simulation studies at the given reference values of energy, integrated luminosity and polarisation. 
For invisible decays of the Higgs, the number quoted is the 95\% confidence upper limit on the branching ratio.
}
\label{tab:stimesbrbase}
  \end{center}
\end{table}

\clearpage

%% file: Appendix-projected.tex

\section{Table of ILC projected uncertainties}

In this appendix, we list the current projections from  the ILC detector
groups for the expected accuracy with which the most important
physics parameters constrained by the ILC will be measured.  
These numbers have been assembled by the LCC Physics Working Group~\cite{LCCPhysGroup:2015},
based on scenario H-20. We
recommend that these numbers be used in discussions of the ILC physics
prospects and in comparisons of the ILC with other proposed
facilities.


Table~\ref{tab:resultsone} gives the corresponding projections for
the uncertainties in physics parameters.   The numbers listed are
obtained
 from full-simulation analyses using
the ILD and SiD detector models, described in~\cite{ILCTDRdetectors}.   For
each number, we have given a reference in which the method is 
described.  The actual number given may reflect more recent
improvements in the analysis.   The uncertainties include both
expected statistical and systematic errors.

The estimated uncertainties for the Higgs boson couplings are based on
a model-independent fit to ILC observables in which all individual couplings (including
the loop-induced couplings to $gg$ and $\gamma\gamma$) are varied
independently.  The total width of the Higgs boson is also taken as an
independent variable, to account for exotic Higgs decays not
constrained by any direct measurement.  (Higgs decay to invisible
states is directly observed using the $hZ$ production process.) 
  The constraints on the Higgs boson derived from
this analysis are completely model-independent.  

The second line for $g(h\gamma\gamma)$ assumes that the ILC data are
combined with an LHC measurement of the ratio of branching ratios 
$\Gamma(h\to \gamma\gamma)/\Gamma(h\to ZZ ^*)$ to 2\% accuracy, as
projected by the ATLAS collaboration for the High-Luminosity
LHC~\cite{ATLAS}.
This is the only place where combination with projected LHC results
significantly improves the model-independent ILC results.

For comparison with results from hadron colliders, where a
model-independent analysis is not possible, the 2013 Snowmass study~\cite{Dawson:2013bba}
suggested
a fit to observables in which one adds the model assumptions that 
$g(hc\bar c)/g(ht\bar t)$ and   $g(h\mu\mu)/g(h \tau\tau)$ have
their Standard Model values and that the Higgs boson has no invisible or
exotic decays.   The results of that analysis are given in
Table~\ref{tab:resultstwo}. 

 Results from the analysis of measurements of the top quark threshold
are dominated by theoretical systematic errors from the calculation of
the threshold cross section shape.  These errors are estimated
conservatively based on a new calculation of this cross section at the 
N$^3$LO level~\cite{Beneke}.   Measurements at any $\ee$ collider should show these 
same uncertainties.  

The estimated uncertainties for the top quark couplings are analyzed
in the following way~\cite{Roman}: First a fit is done with the four
chirality-conserving couplings
$g_L^\gamma, g_R^\gamma, g_L^Z, g_R^Z$ taken to be independent
parameters and the chirality flip couplings taken to be zero.  Then a
fit is done with the chirality conserving parameters taken at their
Standard Model   values and the two  CP-conserving chirality-flip
parameters
$F_2^\gamma$, $F_2^Z$ taken as independent free parameters.    For the
high-luminosity estimates, we have conservatively added a 0.5\%
systematic uncertainty.

The limits on dark matter production are based on the same effective field
theory analysis~\cite{Maxim} as used in section~\ref{subsec:physWIMP}. Dark matter pair production
is represented by a contact interaction with the scale $\Lambda$; the labels D5 and D8 refer to
two possible spin structures.  We emphasize that the
effective field theory approximation is accurate  in this analysis, while
it is questionable in similar analyses for hadron colliders.   The
quoted limits are based on a full-simulation study described in Ref.~\cite{Bartels:2012ex}. 

\begin{table}[h] \begin{center}
\begin{tabular}{lc|c|c|lr}
Topic          &  Parameter   & Initial Phase & Full Data Set &  units
&  ref.  
\\  \hline 
Higgs          &   $ m_h $      &   25    & 15  &  MeV & \cite{Li}
\\
                    &   $   g(hZZ)   $    
                    &   0.58   &  0.31  &   \% &
                    \cite{Asner:2013psa} \\
                   &     $   g(hWW)     $ &  
                   0.81 &  0.42       &    \% &
                    \cite{Asner:2013psa} \\
                   &     $   g(hb\bar b)   $ & 
                  1.5 &  0.7   &    \% &
                    \cite{Asner:2013psa} \\  
                   &    $    g(h g g)   $ & 
                  2.3 & 1.0     &      \% &
                    \cite{Asner:2013psa} \\
                   &    $    g(h \gamma \gamma)   $  &  7.8  &   3.4 &\% &
                    \cite{Asner:2013psa} \\
                 &   &   1.2  & 1.0&   \%,
                  w. LHC results & \cite{MEPHiggs}\\
                   &      $ g(h \tau\tau)   $ & 
                   1.9 &   0.9 &  \% &
                    \cite{Asner:2013psa} \\
                   &    $   g(h c\bar c)    $   &   
                2.7 &   1.2 & \% &
                    \cite{Asner:2013psa} \\
                   &    $   g(h t\bar t)    $  &  18 &   6.3     & \%,
                direct     &
                    \cite{Asner:2013psa} \\
                 &    &  20 &   20     & \%, 
                 $t\bar t$ threshold  &
                    \cite{Horiguchi} \\
                 &    $   g(h\mu\mu)    $   &     20    &  9.2    &  \% &
                    \cite{Asner:2013psa} \\
                   &      $ g(hhh)     $         & 77 &  27     & \% &
                    \cite{Asner:2013psa} \\
                    & $ \Gamma_{tot}$  & 3.8 &
                 1.8  & \% &
                    \cite{Asner:2013psa} \\
                   & $ \Gamma_{invis}$ & 0.54&
                 0.29    & \%,   95\% conf. limit  &
                    \cite{Asner:2013psa} \\
\hline
Top         &    $  m_t $        &  50 & 50  &  MeV ($m_t$(1S))  &
                                   \cite{Seidel:2013sqa} \\
        & $ \Gamma_t $  &60   & 60  &  MeV &  \cite{Horiguchi}   \\ 
                &    $ g_L^\gamma   $ & 0.8  &
               0.6  & \% & \cite{Roman}\\
                          &    $ g_R^\gamma   $ & 0.8  &
               0.6  & \% & \cite{Roman}\\
    &    $ g_L^Z   $ & 1.0 & 0.6  & \% &  \cite{Roman}\\
                &    $  g_R^Z   $ & 2.5 &
               1.0   & \% &  \cite{Roman}\\
                 &  $ F_2^\gamma$  & 0.001 & 0.001  & absolute  &  \cite{Roman}\\
                 &  $ F_2^Z$  &  0.002  & 0.002  &absolute &  \cite{Roman}\\
       \hline
$W$       &  $ m_W   $   &     2.8     &   2.4  & MeV   &\cite{Baak:2013fwa} \\
         &   $ g^Z_1     $    &        $    8.5\times 10^{-4}     $   &
        $  6\times 10^{-4}  $ &   absolute  & \cite{List} \\
  &  $\kappa_{\gamma}   $   &   $  9.2\times 10^{-4}     $    &    $   7 \times 10^{-4}  $   &
  absolute &  \cite{List} \\ 
   &  $\lambda_{\gamma}    $  &  $ 7 \times 10^{-4}    $    &    $   2.5 \times 10^{-4}   $&  absolute &  \cite{List} \\ 
\hline
Dark Matter     &  EFT $\Lambda$: D5  &  2.3 & 3.0 & TeV,
90\% conf. limit  & \cite{Bartels:2012ex}\\ 
&  EFT $\Lambda$:  D8  &  2.2 & 2.8 & TeV,
90\% conf. limit  & \cite{Bartels:2012ex} \\ \hline
\end{tabular}

\caption{Projected accuracies 
 of measurements of Standard Model
  parameters  at the two stages of the ILC program proposed in this
  report.  This table was assembled by the LCC Physics Working Group~\cite{LCCPhysGroup:2015}.
  This program has  an
  initial phase with 500~fb$^{-1}$
  at 500~GeV,   200~fb$^{-1}$  at 350~GeV,  and  500~fb$^{-1}$
  at 250~GeV,  and a luminosity-upgraded phase with an additional 3500~fb$^{-1}$
  at 500~GeV and 1500~fb$^{-1}$
  at 250~GeV.  Initial state polarizations are taken according to the
  prescriptions above.  Uncertainties are
  listed as $1\sigma$ errors (except where indicated),
  computed cumulatively at each stage of the program.  These estimated
  errors include
  both statistical uncertainties and theoretical and experimental systematic
  uncertainties. Except where indicated, errors in 
  percent (\%)  are fractional uncertainties
  relative to the Standard Model values. More specific information for
  the sets of measurements is given in the text. For each measurement, a
  reference describing the technique is given. }
\label{tab:resultsone}
\end{center}
\end{table}

\begin{table}[h] \begin{center}
\begin{tabular}{lc|c|c|l}
Topic          &  Parameter   & Initial Phase & Full Data Set &
\\  \hline 
Higgs          &   $   g(hZZ)   $    
                    &   0.37   &  0.2  &   \% \\
                     &     $   g(hWW)     $ &  
                   0.51 &  0.24      &    \% \\
                   &     $   g(hb\bar b)   $ & 
                  1.1&  0.49   &    \% \\  
                   &    $    g(h g g)   $ & 
                  2.1 & 0.95      &      \% \\
                   &    $    g(h \gamma \gamma)   $  & 7.7  &   3.4 &\% \\
                  &      $ g(h \tau\tau)  , g(\mu\mu) $ & 
                   1.5 &   0.73 &  \%\\
                   &    $   g(h c\bar c), g(ht\bar t)    $   &   
                2.5 &   1.1  & \% \\
                    & $ \Gamma_{tot}$  & 1.8  &
                  0.96  & \% \\   \hline
\end{tabular}

\caption{Projected accuracies of measurements of Higgs boson couplings
  at the two  stages of the ILC program, using the
  model-dependent fit suggested for the Snowmass 2013
  study~\cite{Dawson:2013bba}. The analysis is as described in
  \cite{Asner:2013psa}.  The ILC run plan assumed is the same as in 
  Table~\ref{tab:resultsone}. }
\label{tab:resultstwo}
\end{center}
\end{table}

%% file: bib.tex
\bibliographystyle{apsrev}